\documentclass[10pt]{article}

\usepackage{amsmath}

\usepackage{array}

\usepackage{appendix}

\usepackage{tocloft}                   

\usepackage{graphicx}

\usepackage{amsfonts}

\usepackage{amssymb}

\usepackage{mathrsfs}

\usepackage{yfonts}

\usepackage{euscript}

\usepackage{centernot}                 

\usepackage{ifsym}                     

\usepackage{lmodern}                   

\usepackage{upgreek}

\usepackage{mathtools}

\usepackage{bm}

\usepackage{undertilde}

\usepackage{amsfonts}

\usepackage{amssymb}

\usepackage{mathrsfs}

\usepackage{upgreek}

\usepackage{mathtools}

\usepackage{color}

\usepackage{slantsc}
\usepackage{calligra}

\usepackage{bbold}          

\usepackage[T1]{fontenc}

\usepackage{epsf}

\usepackage{latexsym}

\usepackage{tipa}

\usepackage{makeidx}

\makeindex

\def\b{\begin{equation}}

\def\e{\begin{equation}}

\def\be{\begin{equation}}              

\def\ee{\end{equation}}

\def\beq{\begin{equation}}

\def\eeq{\end{equation}}

\def\bea{\begin{eqnarray}}

\def\eea{\end{eqnarray}}

\def\half{\mbox{$\frac{1}{2}$}}

\def\m{\mbox{ }}

\def\mma {\m , \m \m }

\def\!{\hspace{-1.6667em}}

\def\n{\noindent}

\def\f{\footnote}

\def\u{\underline}

\def\slGamma{\mathit{\Gamma}}                     

\def\slLambda{\mathit{\Lambda}}                   

\def\slOmega{\mathit{\Omega}}                      
\def\bic{\mbox{\boldmath$c$}}

\def\bih{\mbox{\boldmath$h$}}

\def\bil{\mbox{\boldmath$l$}}

\def\bim{\mbox{\boldmath$m$}}

\def\bin{\mbox{\boldmath$n$}}

\def\bip{\mbox{\boldmath$p$}}

\def\biq{\mbox{\boldmath$q$}}

\def\biv{\mbox{\boldmath$v$}}

\def\biy{\mbox{\boldmath$y$}}

\def\biz{\mbox{\boldmath$z$}}

\def\biM{\mbox{\boldmath$M$}}

\def\biN{\mbox{\boldmath$N$}}

\def\biP{\mbox{\boldmath$P$}}

\def\biQ{\mbox{\boldmath$Q$}}

\def\sbin{\mbox{\scriptsize\boldmath$n$}}

\def\sbiM{\mbox{\scriptsize\boldmath$M$}}

\def\sbiN{\mbox{\scriptsize\boldmath$N$}}

\def\bbeta{\mbox{\boldmath$\beta$}}

\def\bslLambda{\mbox{\boldmath$\Lambda$}}         

\def\me{\mbox{e}}

\def\mg{\mbox{g}}

\def\mh{\mbox{h}}

\def\mt{\mbox{t}}

\def\bh{\u{\u{\mbox{h}}}  }            

\def\bg{\mbox{\bf g}}

\def\bh{\mbox{\bf h}}

\def\bq{\mbox{\bf q}}

\def\bupSigma{\mbox{\boldmath$\Sigma$}}                 

\def\fA{\mbox{\sffamily A}}           

\def\fC{\mbox{\sffamily C}}

\def\fP{\mbox{\sffamily P}}

\def\fS{\mbox{\sffamily S}}

\def\sa{\mbox{\scriptsize a}}

\def\sb{\mbox{\scriptsize b}}

\def\scc{\mbox{\scriptsize c}}

\def\sd{\mbox{\scriptsize d}}

\def\se{\mbox{\scriptsize e}}

\def\sm{\mbox{\scriptsize m}}

\def\sn{\mbox{\scriptsize n}} 

\def\so{\mbox{\scriptsize o}}

\def\sr{\mbox{\scriptsize r}}

\def\sss{\mbox{\scriptsize s}}  

\def\st{\mbox{\scriptsize t}}

\def\sw{\mbox{\scriptsize w}}

\def\sB{\mbox{\scriptsize B}}

\def\sE{\mbox{\scriptsize E}}

\def\sG{\mbox{\scriptsize G}}

\def\sI{\mbox{\scriptsize I}}

\def\sJ{\mbox{\scriptsize J}}

\def\sL{\mbox{\scriptsize L}} 

\def\sM{\mbox{\scriptsize M}} 

\def\sN{\mbox{\scriptsize N}}

\def\sP{\mbox{\scriptsize P}}

\def\sR{\mbox{\scriptsize R}}

\def\sS{\mbox{\scriptsize S}}

\def\sT{\mbox{\scriptsize T}}

\def\sfA{\mbox{\sffamily{\scriptsize A}}}     

\def\sfB{\mbox{\sffamily{\scriptsize B}}}     

\def\sfC{\mbox{\sffamily{\scriptsize C}}}     

\def\sbcC{\mbox{\boldmath \scriptsize ${\cal C}$}}

\def\bscP{\mbox{\boldmath\scriptsize${\cal P}$}}                               

\def\bscS{\mbox{\boldmath \scriptsize${\cal S}$}}                               

\def\sumi2{\sum\mbox{}_{\mbox{}_{\mbox{\scriptsize $i$=1}}}^2}

\def\sumi3{\sum\mbox{}_{\mbox{}_{\mbox{\scriptsize $i$=1}}}^3}

\def\sumABcycles3{\sum\mbox{}_{\mbox{}_{\mbox{\scriptsize cycles $A,B$=1}}}^{3}}

\def\sumCDcycles3{\sum\mbox{}_{\mbox{}_{\mbox{\scriptsize cycles $C,D$=1}}}^{3}}

\def\sumj3{\sum\mbox{}_{\mbox{}_{\mbox{\scriptsize $j$=1}}}^3}

\def\sumk3{\sum\mbox{}_{\mbox{}_{\mbox{\scriptsize $k$=1}}}^3}

\def\prodiA1{\prod\mbox{}_{\mbox{}_{\mbox{\scriptsize $i$=1}}}^{A - 1}}

\def\bigtimes{\mbox{\Large $\times$}}

\def\d{\textrm{d}}                                                  

\def\pa{\partial}                                                   

\def\Last{\mbox{\Large$\ast$}}                                      

\def\Ast{\mbox{\Large$\ast$}}                                       %

\def\es{\m = \m}

\def\:={\m := \m}

\def\=:{\m =: \m}

\def\FrT{\mathfrak{T}}                                         

\def\lFrs{\mathfrak{S}}                                        

\def\FrU{\mbox{$\mathfrak{U}$}}                                
\def\Frm{\mbox{\Large $\mathfrak{m}$}}                         
\def\FrM{\mbox{$\mathfrak{M}$}}                                
\def\lFrg{\mbox{\Large$\mathfrak{g}$}}                         
\def\FrT{\mbox{\boldmath$\mathfrak{T}$}}                       
\def\Hilb{\mbox{{\boldmath$\mathfrak{H}$}ilb}}                 
\def\lt{\mbox{\Large $t$}}                                 

\def\scC{\mbox{\scriptsize ${\cal C}$}}                    
\def\scE{\mbox{\scriptsize ${\cal E}$}}                    

\def\scH{\mbox{\scriptsize ${\cal H}$}}                    

\def\bscP{\mbox{\boldmath\scriptsize ${\cal P}$}}

\def\bscP{\mbox{\boldmath\scriptsize ${\cal P}$}}

\def\bscS{\mbox{\boldmath\scriptsize ${\cal S}$}}

\def\Chronos{\scC\mbox{hronos}}                            

\def\FrQ{\mbox{\Large $\mathfrak{q}$}}                               
\def\sFrQ{\mbox{\large $\mathfrak{q}$}}                              
\def\Phase{\mbox{{\boldmath$\mathfrak{P}$}hase}}                     

\def\bFrR{\mbox{\boldmath$\mathfrak{R}$}}                            
\def\Rig-Phase{\bFrR\mbox{ig-}\Phase}                                
                                                                													   
\def\bFrR{\mbox{\boldmath$\mathfrak{R}$}}                            

\def\bFrR{\mbox{\boldmath$\mathfrak{R}$}}                            

\def\1mat{\u{\u{1}}}                                                 

\def\Mini{\mbox{{\boldmath$\mathfrak{M}$}ini}}                       
\def\Ani{\mbox{{\Large$\mathfrak{a}$}ni}}                   

\def\Positive-Modespace{\mbox{{\boldmath$\mathfrak{M}$}odespace$^+$}}

\def\POSITIVE-MODESPACE{\mbox{{\boldmath$\mathfrak{M}$}ODESPACE$^+$}}
                                                                                                                             														
\def\Riem{\bFrR\mbox{iem}}                                           

\def\Kin-Hilb{\mbox{{\boldmath$\mathfrak{K}$}in-\Hilb}}                     

\def\Mid-Hilb{\mbox{{\boldmath$\mathfrak{M}$}id-\Hilb}}                     

\def\Dyn-Hilb{\mbox{{\boldmath$\mathfrak{D}$}yn-\Hilb}}                     

\def\5Star{\mbox{\Large$\star$}}              

\begin{document}

\begin{center}

\Huge{\bf A LOCAL RESOLUTION OF}

\Huge{\bf THE PROBLEM OF TIME}

\vspace{.1in}

\Large{\bf I. Introduction and Temporal Relationalism}

\vspace{0.05in}

{\bf  E.  Anderson}$^1$ 

{\bf\it based on calculations done at Peterhouse, Cambridge} 

\end{center}

\begin{abstract}

This Series of Articles provides a local resolution of this major longstanding foundational problem between QM and GR, 
or, more generally, between Background Dependent and Background Independent Physics.  
We focus on the classical version; the concepts we use are moreover universal enough to admit quantum counterparts. 
This requires a series of articles to lay out because the Problem of Time is multi-faceted and is largely about interferences between facets, 
with traditional solutions to individual facets breaking down in attempted joint resolutions of facets.  
\cite{ABook} already covered this at both the classical and semiclassical quantum levels. 
This Series serves, firstly, to isolate this resolution from \cite{ABook}'s introductory and field-wide comparative material, 
passing from an 84-part work down to just a 14-part one.  
Secondly, to clarify that, at the usual differential-geometric level of structure used in physical theories, 
our classical local resolution is entirely catered for by Lie's mathematics. 
This renders our local classical part of the Problem of Time understandable by a very large proportion of Theoretical Physics or Mathematics majors.

In this first article, we cover a first aspect: Temporal Relationalism. 
This generalizes the facet traditionally known as the Frozen Formalism Problem, that follows from the quantum form of GR's Hamiltonian constraint.    

\end{abstract}

$^1$ dr.e.anderson.maths.physics *at* protonmail.com
			  
\section{Introduction}
				 
\subsection{The Problem of Time and underlying Theory of Background Independence}

Time is conceived of substantially differently across the observationally established paradigms of Physics.
Detailed consideration of this -- the subject of Part I of \cite{ABook} -- reveals the main chasm to be between 
Newtonian Physics,  
Special Relativity (SR),
Quantum Mechanics (QM), 
Quantum Field Theory (QFT) on the one side, and 
General Relativity (GR) on the other.  

\m 

\n This chasm is present more generally between Background Dependent paradigms -- such those of as Newtonian Physics, SR, QM and QFT on the one hand, 
and Background Independent paradigms \cite{A64, A67, Giu06, ABook}, such as GR's, on the other. 
[The first four do of course differ among themselves to some extent in temporal and spatial conceptualization, 
but to a much lesser extent than any of them differ from GR in these regards.] 
This chasm is thus a {\sl Paradigm Split}, and can moreover be envisaged as 
the modern form taken by the Absolute versus Relational (Motion) Debate \cite{DoD, Buckets, ABook}. 
This dates at least as far back to Newton \cite{Newton} and his absolute conceptions space and time versus Leibniz and Mach's relational objections \cite{L, M}.  

\m 

\n The Problem of Time is moreover multi-faceted, because there are multiple differences in temporal and spatial conceptualization across this chasm, 
and some of these having further consequences for the nature and form of Physical Law.  
Most of its facets were already known to Wheeler \cite{WheelerGRT, Battelle} and DeWitt \cite{DeWitt67} in the 1960s or to Dirac \cite{DiracObs, Dirac, Dirac51, Dirac58} in the 1950s.
Many authors have since attempted to resolve the Problem of Time. 
See e.g. \cite{K81, K91} for early reviews, 
\cite{K92, I93} for Kucha\v{r} and Isham's seminal progress,  
\cite{APoT} for a summary thereof, 
Part I of \cite{ABook} for grounding the Problem of Time on established Physics' differing notions of time, 
\cite{APoT2, APoT3, A-Lett} and Parts II and III of \cite{ABook} for further progress, 
and the Appendix Part of \cite{ABook} for a suitable course in supporting mathematical techniques for understanding this literature.  

\m 

\n Kucha\v{r} and Isham's seminal progress \cite{K92, I93} consists firstly of formalizing the conceptual classification into Problem of Time facets, 
giving 8 such (Fig \ref{1}.a).  
Secondly, in showing how every attempt to solve the Problem of Time up to that point fails when examined in sufficient detail. 
While some works were shown to fail to overcome even one facet, most were moreover shown to break down upon 
attempting to combine resolutions of piecemeal facets to jointly resolve facets. 
I.e.\, in addition to having multiple facets, the lion's share of the Problem of Time consists of interferences between facets.  
Such interferences occur, moreover, because of the various facets having common origins in the chasm between Background Dependent and Background Independent Physics.  
For sure, the current Series' titular `Problem of Time' refers to multi-faceted and facet-interfering conceptualizations along such lines. 
This is to be contrasted with works that confuse the Problem of Time with just one of its facets, thus missing out on all the other facets 
and thus also furthermore on the lion's share of the Problem of Time: the interferences between the multiple facets.  
While such works are of clearly much more likely to solve what they purport to be 'the Problem of Time', 
this is of course very likely to just be adding to the literature of single-facet resolutions which fail to combine into joint resolutions of all Problem of Time facets. 

\m

\n Whereas there has been quite widespread belief that the Problem of Time is a quantum matter, this is in fact a further misconception.
Emphasizing the underlying Background Dependence versus Independence chasm renders this clear, 
since Newtonian Physics and SR on the one hand versus GR on the other already exhibits almost all Problem of Time facets.
This has the benefit of the simpler classical version providing a model arena for the more complicated quantum version.  
This benefit is by using the mathematically simpler and thus more readily surmountable classical Problem of Time facets as a testing ground for structures and methods, 
some of which then transcend to the quantum level, whether directly or via suggesting harder quantum counterparts.  
The current Series serves to improve on this classical approach, by further identifying the mathematics in use and demonstrating the sufficiency of Lie mathematics in this regard 
at the classical level for the usual differential-geometric rendition of the classical laws of Physics.

\m 

\n The Author's previous further progress consisted in, 
firstly, incorporating spacetime-primary approaches into the conceptual classification to compensate for Isham and Kucha\v{r}'s classification's canonical-primality bias.  
See Part I of \cite{ABook} for detailed motivation of each of these primalities.  

\m 

\n Secondly, the passage from these Problem of Time facets to a classification 
of an equal multiplicity of underlying theory-independent classical-or-quantum Background Independence aspects.
This amounts a renaming of facets to more truly reflect each's conceptual content.  
The chain of thought involved is spread out over Articles I to IV, 
culminating in IV's Conclusion's summary figure of the passage to Fig \ref{1}.b) from Fig \ref{1}.a).

\subsection{Some underlying principles} 

A starting point for considerations of Background Independence is as follows.   

\m 

\n{\bf Relationalism-0)} {\sl Physics is to solely concern relations between tangible entities}.\footnote{These are not `just matter', 
and are named thus, via Isham, along the lines of Heidegger.}  

\m 

\n Some key diagnostics of `tangible entities' are as follows.   

\m

\n{\bf Relationalism-1)} Tangible entities {\sl act testably and are actable upon}. 
%

\m

\n Things which do not act testably or cannot be acted upon are held to be {\sl physical} non-entities.
These can still be held to be a type of thing as regards being able to {\sl philosophize} about them or {\sl mathematically represent} them.  
Absolute space is an obvious archetypal example of such a non-entity.  
Relational intuition is that imperceptible objects should not be playing causal roles influencing the motions of actual bodies. 
As a first sharpening of this, James L. Anderson's \cite{A67} stated that 
{\it ``the dynamical quantities depend on the absolute elements but not vice versa"}, 
and an absolute object {\it ``affects the behavior of other objects but is not affected by these objects in turn"} \cite{AG}. 

\m

\n{\bf Relationalism 2) [Leibniz's Identity of Indiscernibles]} \cite{L} {\sl Any entities indiscernible from each other are held to be identical}.  

\m

\n{\bf Remark 3} This posits that physical indiscernibility trumps multiplicity of mathematical representation.  
Such multiplicity still exists mathematically, but the mathematics corresponding to the {\sl true} physics in question is the equivalence class spanning that multiplicity.
One would only wish to attribute physical significance to calculations of tangible entities which are independent of the choice of representative of the equivalence class. 
By this e.g.\ our Universe and a copy in which all material objects are collectively displaced by a fixed distance surely share all observable properties, 
and so they are one and the same.  
An archetype of such an approach in modern Physics is Gauge Theory (see Articles IV and VII).   
This additionally factors in the major insight that a mixture of tangible entities and non-entities is often far more straightforward to represent mathematically.  

\m

\n{\bf Remark 4} For now, consider separate treatments of time on the one hand, and space, configurations, dynamics and canonical formulation on the other.  
This befits the great conceptual heterogeneity between these (see Part I of \cite{ABook} for details).  
Once this is understood, relational postulates can be stated, and a coherent subset of these are sharply mathematically implementable. 
This leads to rejecting absolute time and absolute space, and, more eventually (as argued in Part I of \cite{ABook}) to the first 2 Background Independence aspects.

\subsection{Temporal and Configurational Relationalisms}\label{TCR}

\n {\bf Temporal Relationalism} (aspect 0a) \cite{L, FileR} is that {\sl there is no meaningful time for the Universe as a whole at the primary level}.  

\m 

\n We shall see that this is implemented by actions \cite{Jacobi, Synge, BSW, Magic, BB82, B94I, FORD, FileR, ABook} which are, 
firstly,  free of extraneous time-like quantities, and, 
secondly, with no physically meaningful role for `label times' either (strategy 0a-1).  

\m

\n Temporal Relationalism leads to the notorious {\bf Frozen Formalism Problem} \cite{Battelle, K92, I93} (facet 0a). 
This is more widely known at the quantum level, where an apparently frozen quantum wave equation -- the Wheeler--DeWitt equation \cite{Battelle, DeWitt67} -- 
occurs in a context in which one would expect an equation which is dependent on (some notion of) time.
This is moreover unfortunately often confused with the entirety of the multi-faceted Problem of Time.

\m 

\n The current Series' approach to frozenness is to recover time at the secondary level from Mach's `time is to be abstracted from change', 
i.e.\ an {\bf emergent Machian time strategy} (strategy 0a-2).  

\m

\n If time is not primary, moreover, we need to study whatever other entities that are still regarded as primary; 
this starts with configurations $\biQ$ and configuration spaces $\FrQ$.    

\m

\n {\bf Configurational Relationalism} (aspect 0b) \cite{Cauchy, Burnside, BSW, WheelerGRT, BB82, Kendall84, FORD, FileR, PE16, ABook, S-III, Minimal-N, A-Killing, 
A-Cpct} involves taking into account that a continuous group of transformations $\lFrg$ acting on the system's configuration space $\FrQ$ is physically irrelevant.   
For Mechanics, these transformations are usually translations and rotations of space, 
though in general Configurational Relationalism also covers physically irrelevant internal transformations, as occur in the most common types of Gauge Theory.  

\m 

\n Configurational Relationalism can be implemented, at least in principle, by {\bf Best Matching} (strategy 0b), 
i.e.\ is bringing two configurations into minimum incongruence with each other by application of $\lFrg$'s group action. 
In the case of GR -- for which $\lFrg = Diff(\bupSigma)$: the spatial diffeomorphisms -- Configurational Relationalism leads to 
the {\bf Thin Sandwich Problem} (facet 0b). 
This is a particular GR specialization of the abovementioned notion of Best Matching.  
A more general strategy for Configurational Relationalism is the $\lFrg${\bf -Act} $\lFrg${\bf -All Method}: 
group action followed by involvement of all the group, for which Article II argues that the widely known group averaging is a useful prototype. 

\m 

\n Temporal and Configurational elationalism give two precise senses in which GR is `Machian' \cite{M, ABook}: 
Mach's Time Principle (Sec \ref{MTP}) and Mach's Space Principle (Sec II.5.1).  
Mach's work is widely of foundational interest; for instance, some of Mach's concepts played a role in Einstein's search for GR.
The above two Machian attributes do not coincide with how Einstein interpreted a partly different set of Mach's ideas; 
his historical route to GR ended up making at best indirect use of Machian themes.  
As Wheeler argued \cite{Battelle, MTW}, however, there are many routes to GR. 
Some of these routes arrive at a dynamical formulation of GR: a theory of evolving spatial geometry: Geometrodynamics \cite{Battelle}.
It then turns out that a more specific formulation of GR-as-Geometrodynamics is Machian after all (see \cite{RWR, AM13} and Articles II and VI).
Finally, GR in Machian Geometrodynamics form can furthermore 
even be rederived from Temporal and Configurational Relationalism first principles (see \cite{RWR, AM13} and Article IX).

\m 

\n Each of Temporal and Configurational Relationalism moreover provides constraint equations.
In the case of GR, these are, respectively, 
the well-known Hamiltonian and momentum constraints that usually occupy centre-stage in accounts of the Problem of Time.   
Indeed, the abovementioned Wheeler--DeWitt equation is the quantum Hamiltonian constraint, 
whereas the Thin Sandwich Problem is a particular approach to solving the momentum constraint at the classical level.

\subsection{Model arenas} 

Aside from the classical simplification, this Series of Articles makes use of the simpler {\bf Relational Particle Mechanics (RPMs)} and 
                                                                                          {\bf Minisuperspace} model arenas 
prior to passing to more complicated cases; these arenas are introduced in Secs II.4.1 and \ref{MSS-Intro} respectively.  

\m 

\n{\bf Diffeomorphisms} are, moreover, crucial \cite{I93} as regards a number of Problem of Time facets; 
to feature nontrivially, 

\n these require as a minimum inhomogeneous GR models.  

\m 

\n Balancing this requirement, enough simplicity for calculations, and cosmological applications, 
this Series of Articles' third choice of model arena is {\bf Slightly Inhomogeneous Cosmology (SIC)}: 
a type of perturbative Midisuperspace model; see Article XI for more.  
This furthermore permits investigating whether galaxies and cosmic microwave background hot-spots 
could have originated from quantum cosmological fluctuations \cite{HallHaw}.  
Finally, this choice of model arenas amounts to concentrating on Quantum Cosmology rather than Black Hole models.

\subsection{Constraints and canonical observables}

Following on from Sec \ref{TCR}, it is next natural to ask whether one has found all of the constraints. 
I.e. {\bf Constraint Closure} (aspect 3) in an algebraic sense. 

\m 

\n If the answer is in the negative, one has a {\bf Constraint Closure Problem} (facet 1) \cite{K92, I93, APoT2, APoT3, ABook}.

\m 

\n This is approached by introducing a suitable brackets structure and systematically applying the {\bf Dirac Algorithm} (strategy 1) \cite{Dirac, HTBook}.  

\m 

\n This consistency established, we show it subsequently makes sense to consider which objects brackets-commute with the constraints, 
or with specific subalgebraic structures thereof.  
These objects --    {\it observables} \cite{DiracObs, HTBook, K92, I93, AObs, AObs2, AObs3, ABook, DO-1} -- are useful objects due to their physical content, 
whereby aspect 2 is {\bf Assignment of Observables}.

\m 

\n If obtaining a sufficient set of these to do Physics is in practice blocked -- a common occurrence in Gravitational Theory -- 
then one has a {\bf Problem of Observables} (facet 2) \cite{K92, I93, AObs, ABook}.   

\m 

\n Observables can moreover already be defined in the absense of constraints. 
Finding observables, with or without constraints applying, amounts to, given a state space (for now phase space) 
{\bf Finding Function Spaces Thereover} (strategy 2) \cite{ABook}.

\subsection{Spacetime Constructability}

Starting with less structure than spacetime -- assuming just one or both of spatial structure or less levels of mathematical structure for 
spacetime -- is particularly motivated by Quantum Theory \cite{Battelle}. 
Space from less mathematical levels of structure of space is also a valid pursuit. 
In such approaches, the spacetime concept is moreover to hold in suitable limiting regimes: {\bf Spacetime Constructability} is required.

\m 

\n Let us for now focus on {\bf Spacetime Constructability from Space} (aspect A).

\m 

\n If this is false, or remains unproven, then we have a {\bf Spacetime Construction Problem (from Space} (facet A).

\m 

\n A classical-level Spacetime Construction was given in \cite{RWR, Phan} 
and properly decked out to comply with the other local Problem of Time facets in \cite{AM13, ABook}.  
Strategy A is {\bf Feeding Families of Theories into the Dirac Algorithm}. 
This can be interpreted as {\bf deforming algebraic structures}, and it works out in cases for which {\bf Rigidity} manifests itself.
We reformulate this in terms of deformations of algebraic structures.  

\m 

\n We refer to {\bf Space Constructability from less Space Structure} as aspect 3 of Background Independence; 
this pursuit being free from time, it is not counted among the Problem of Time facets.  
We shall see that the strategy for this parallels the preceding one while making use of the more general Lie Algorithm.

\subsection{Aspects concerning spacetime}

Since GR is also a theory with a meaningful and nontrivial notion of spacetime, it has more Background Independence aspects than Relational Particle Mechanics does. 
Indeed, the Einstein field equations of GR determine the form of GR spacetime, as opposed to SR Physics unfolding on a fixed background spacetime.  
From a dynamical perspective, GR's geometrodynamical evolution {\sl forms spacetime itself}, 
rather than being a theory of the evolution of other fields {\sl on} spacetime or {\sl on} a sequence of fixed background spatial geometries.  
Regardless of whether spacetime is primary or emergent, there is now also need for the following.

\m 

\n The current Article serves, firstly, to further subdivide `spacetime' Background Independence \cite{APoT2, ABook} into the following.  

\m 

\n{\bf Spacetime Relationalism}                                  (aspect 0$^{\prime}$). 

\m 

\n{\bf Spacetime Generator Closure}                              (aspect 1$^{\prime}$).         

\m 

\n{\bf Assignment of Spacetime Observables}                      (aspect 2$^{\prime}$). 

\m 

\n{\bf Spacetime Constructability from less Spacetime Structure} (aspect 3$^{\prime}$). 

\m 

\n This increases facet-and-aspect count from \cite{APoT3, ABook}'s 9 to 12. 
This four-way partition serves to more closely match conceptualization with the canonical approach.

\m 

\n In Spacetime Relationalism, the {\it diffeomorphisms of spacetime} itself, $Diff(\Frm)$, the physically redundant transformations.  
Whereas this is straightforwardly implemented in the classical spacetime formulation of GR, and these attain Closure as well, 
implementation becomes harder at the quantum level. 
For instance, it feeds into the {\bf Measure Problem} \cite{K92, I93} of Path Integral Approaches to Quantum Gravity.  
The {\bf Problem of Spacetime Observables} (facet $2^{\prime}$) is moreover significant even at the classical level.  
The {\bf Problem of Spacetime Construction from less Spacetime Structure} (facet $3^{\prime}$) 
is a comparative point between theories rather than an essential part of the current basic treatise.  
%

\m 

\n {\it Foliations} of spacetime play major roles, both in dynamical and canonical formulations, 
and as a means of modelling the different possible fleets of observers within approaches in which spacetime is primary.  
Background Independence Physics is moreover to possess {\bf Foliation Independence} (aspect B) \cite{K92, I93}.

\m 

\n If this cannot be established, or fails, then a {\bf Foliation Dependence Problem} is encountered (facet B). 
{\bf Refoliation Invariance} (strategy B) resolves this for GR at the classical level \cite{Tei73, TRiFol, ABook}.

\subsection{Globality, nonuniqueness and `A Local Problem of Time' subproblem}

\n The final facets-and-aspects are as follows.  

\m 

\n{\bf Global Validity}.   

\m 

\n{\bf Unexplained Multiplicities}.  

\m 

\n These are criteria which apply all the other aspects, facets and strategies toward resolving these. 
Contentions with them are termed, respectively, as follows.  

\m 

\n {\bf Global Problems of Time}          \cite{K92, I93} as covered in Epilogues II.B and III.B of \cite{ABook}, \cite{A-CBI} and Article XIV's Conclusion.

\m 

\n {\bf Multiple Choice Problems of Time} \cite{K92, I93, Gotay00}  and Epilogue           III.C of \cite{ABook}.
The main part of this would appear to be the Gronewold--van Hove phenomenon in the classical to quantum bridge.  

\m  

\n All in all, the Problem of Time is a multi-faceted subset of the reasons why forming `Quantum Gravity' Paradigms is difficult and ambiguous; 
further reasons are purely technical, or a mixture of both.
				 
\m  
 
\n 10 facets or 11 aspects remain if these last two are disregarded. 
This moreover remains a consistent problem: seeking for A Local Resolution of the Problem of Time.  
While the Author's abovementioned first two pieces of progress further clarify what the Problem of Time is, 
the Author's third piece of progress is in recently giving an actual local resolution of the Problem of Time. 
See \cite{A-Lett} for a 5-page summary and \cite{ABook} for a 900-page exposition.  
This Series of Articles concentrates on the classical version of {\bf A Local Resolution of the Problem of Time}. 
We set out to streamline Parts I and II of \cite{ABook}, by omitting both how the Problem of Time sits on established Physics' conflicting notions of time 
                                                      and survey of other works exploring alternative strategies for facet resolution. 
We thus now offer bridging 90 and 240-page versions (excluding references): 
respectively, Articles I-IV largely without facet interference, and Articles I-XIV including facet interference.

\subsection{Outline of the rest of this Article} 

We present a first aspect of Background Independence -- Temporal Relationalism -- and the corresponding Problem of Time facet 
(most well-known as the quantum-level Frozen Formalism Problem). 

\m 

\n The first part of Temporal Relationalism is to implement the Leibnizian principle that 
`there is no time for the Universe as a whole at the primary level' (Sec 3) 
This involves using Jacobi's Principle rather than Euler--Lagrange's. 
Combining this resolution with those of all the other facets requires moreover reworking around half of the Principles of Dynamics \cite{Lanczos, Goldstein} 
into Temporally Relational implementing (TRi) form. 
This can be thought of as `taking Jacobi's Principle' more seriously than Jacobi himself did, and thus reworking the rest of Principles of Dynamics 
to follow from this in place of Euler--Lagrange's.  
Sec 3 thus preliminarily presents the parts of the standard Principles of Dynamics \cite{Lanczos, Goldstein} 
that the current Article supplants in the fundamental context of building up A Local Resolution to the Problem of Time.  
This rests on Sec 2's preliminary outline of configuration space $\FrQ$ \cite{Lanczos, DeWitt67, Fischer70, Magic, FM96, Kendall, Giu09, PE16, ABook, S-I, S-II, S-III, Minimal-N}, 
configurations being `what is left' to build upon in primarily timeless pictures, including via building velocity, change or momentum bundles thereover.  

\m 

\n The second part of Temporal Relationalism (Sec 4) is to resolve primary-level timelessness in the manner of Mach's `time is to be abstracted from change'. 
Sec 5 comments further on the consequent classical Machian emergent time.  
%
{            \begin{figure}[!ht]
\centering
\includegraphics[width=1.0\textwidth]{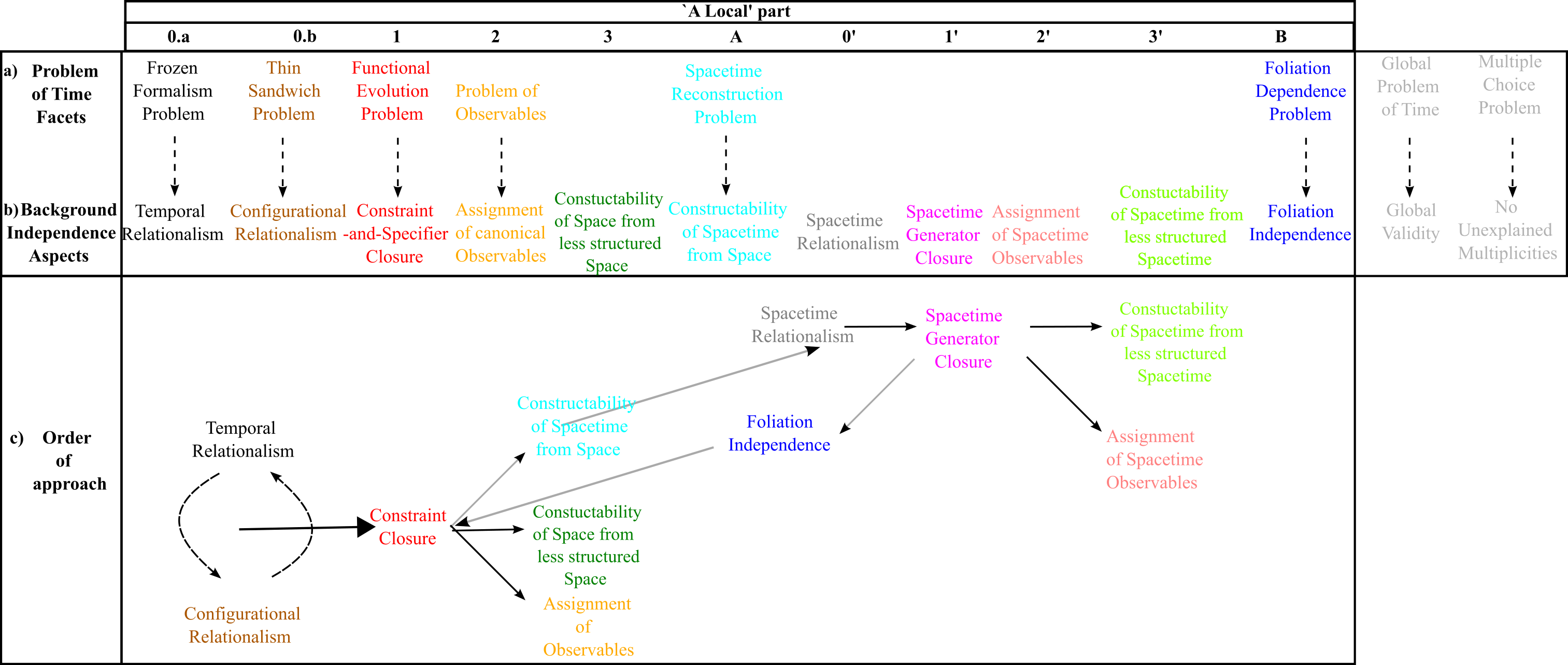}
\caption[Text der im Bilderverzeichnis auftaucht]{\footnotesize{Evolution of conceptualization and nomenclature of 
a) Kucha\v{r} and Isham's \cite{K92, I93} Problem of Time facets into 
b) underlying Background Independence aspects over the course of this Series of Articles. 
This Figure's colour scheme for the first eleven columns is further used in Articles V to XIII's presentation 
of each facet and of how facets interfere with each other.  
12/13ths of these aspects are moreover already classically present: 
all bar the issue of physically and conceptually unaccounted-for multiplicities. 
Solid black arrows connect each primality's parts, while grey arrows form the Wheelerian two-way route between primalities. 

\m 

\n c) The nonlinear order in which the current Series incorporates Background Independence aspects, 
                                    or, equivalently, resolves     Problem of Time         facets.    } }
\label{1}\end{figure}            }

\subsection{Outline of the rest of this Series} 

Article II covers Configurational Relationalism as resolved by the $\lFrg$-act, $\lFrg$-all method. 
\n Article III covers the other 9 classical local Problem of Time facets, 
and Article IV covers all 11 local facets at the quantum level alongside giving a piecemeal-level conclusion. 

\m 

\n Articles V and VI combine Temporal and Configurational Relationalism, for Finite Theories and Field Theories -- including GR -- respectively.  
Article VII unifies classical treatment of Temporal Relationalism, 
                                              Configurational Relationalism and 
											  Constraint Closure, using a suitably TRi Dirac-type Algorithm \cite{ABook}.  										  
Article VIII extends this combination to Assignment of Observables as well. 
Article IX extends Article VII to generalization to instead include classical Spacetime Constructiblility. 
Articles VIII and IX are independent as per the fork in Fig \ref{1}.c).  
Spacetime constructed, Article X considers its own Relationalism, Closure and Assignment of Observables. 

\m 

\n Article XI serves as an arena pit stop, 
introducing Slightly Inhomogenous Cosmology (SIC) \cite{HallHaw, SIC-1, SIC-2, ABook} to cover the minisuperspace and RPM arenas' increasing inadequacy in modelling GR spacetime aspects. 
This Article includes reappraising all aspects covered so far in this further model arena. 
This arena has moreover the added benefit of being along the lines of Halliwell--Hawking's setting for origin of structure of the universe: 
a quantum-level and more Background Independent version of the same kind of model actually used in Observational Cosmology. 
I.e. a likely setting for eventual first observation of semiclassical quantum-gravitational effects.  

\m 

\n Article XII covers Foliation Independence as resolved by Refoliation Invariance. 

\m 

\n Articles V-X and XII involve `mild recategorization' creating a mathematically-consistent formalism for all of a local Problem of Time's facets concurrently.  
This involves in particular having to rewrite around half of the Principles of Dynamics used, in a physically equivalent but now satisfactorily Temporally Relational form. 
We emphasize that the justification for introducing our new Principles of Dynamics is not new problem-solving capacity at the level of numerous small exercises 
but rather that it resolves a 50-year-old fundamental question: the Problem of Time.  
In particular, it is a {\it Temporal Relationalism implementing Principles of Dynamics (TRiPoD)} \cite{TRiPoD, ABook, AM13, MBook}.  
Article XII requires TRiFol as well (foliations); subsequent quantum-level work requires TRiCQT (Canonical Quantum Theory) and TRiPIQT (Path Integral Quantum Theory).  
Article XIII is the series' combined-facets-level Conclusion, whereas Article XIV serve provide the current Series' Lie Mathematics in a self-contained manner.

\m 

\n Another well known joint conceptualization of aspects is viewing Spacetime Construction from Space 
                                                                and Refoliation Invariance   as a Wheelerian two-way route between our two competing primalities. 
Namely spacetime on the one hand versus space, configuration, dynamics or canonical formalism on the other.

\subsection{We just use Lie's mathematics but in a subsequent setting following from QM and GR}

\n While this Series' Problem of Time resolution follows \cite{ABook}, 
we now further identify all the mathematics in the classical version of this as lying within the differential-geometric mathematics of Lie. 
Namely, {\bf Lie derivatives} \cite{Yano55, Yano70}, 
        {\bf Lie brackets} \cite{Serre}, 
		{\bf Lie algebras} \cite{FHBook}, 
		{\bf Lie groups}   \cite{Gilmore, Serre} 
	and the {\bf Flow Method} \cite{John, Olver, Lee2, PE-1, DO-1} for solving PDE systems.   
This is a very useful observation from the points of view of clarity, simplicity, exposition and pedagogy. 
For Lie's Mathematics is widely familiar to physicists, to mathematicians working on continua, and, increasingly, to scientists in other STEM subjects.  
This represents a large improvement in removing both actual and perceived difficulty from a Local Resolution of the Problem of Time 
                                                                                     and its underlying Theory of Background Independence. 
It is now likely that most graduate students in Mathematics, Theoretical Physics, or Applied Differential Geometry-based areas of other STEM subjects can 
now follow the current Series' fundamental developments.    
																					 																					 
\m 

\n The Problem of Time moreover operates in a {\it mathematical amphitheatre} 
that was partly unemphasized and partly undiscovered in the epoque of Lie's own zenith (the 1880's and 90's \cite{Lie}).    
This reflects that the Problem of Time's most severe form requires both QM and GR to be in play.  
QM and GR are 1920's and 1910's developments respectively, 
with GR's dynamics and canonical form having to await the 1950's and 60's \cite{B52, Dirac51, Dirac58, ADM, Dirac}.
In a nutshell, we apply Lie's mathematics to the configuration space and phase space mathematical amphitheatres in Poisson-brackets and Hamiltonian formalisms 
as are jointly suitable for constraints like GR's and passage to QM.  

\m 

\n This ties down prerequisites to understand the current Series to the following.  
Firstly, Lie's approach to Differential Geometry (both widely known and with specifics catered for in Article XIV).  
Secondly, an MA-level course in the Principles of Dynamics, for which \cite{Lanczos} and Chapters 1-2 of \cite{Dirac} are excellent background reading. 
Articles II-IV and VI-XIII also require being familiar with an introductory account of Geometrodynamics, such as \cite{ADM} or Chapter 43 of \cite{MTW}.  

\m

\n Let us furthermore term the above multiply-occurring pieces -- i) Relationalism, ii) Closure, iii) Assignment of Observables -- {\it superaspects}.  
In particular, making careful distinction between assigning canonical observables and spacetime observables 
explains a fair amount of facet ordering difficulties and previous authors misunderstanding each other's work. 
Much of this was `hidden under' using just the word `observables' without sufficient definition and distinction between types of observables, 
whether within authors' own papers or relative to each others' uses 

\m 

\n We will argue that Relationalism, whether Temporal, Configurational or Spacetime, Relationalism is implemented by {\bf Lie derivatives}. 
Closure, whether Constraint or Spacetime Generator, is moreover a matter of {\bf Lie brackets consistency}. 
Constraints themselves form a {\bf Lie algebraic structure}: the constraints algebraic structure.
Next, Assigning Observables, canonical or spacetime, involves {\bf associated Lie algebraic zero-commutant definitions} -- of observables -- realized by further 
{\bf associated Lie algebraic structures}: the observables algebraic structures. 
These refine Taking Function Spaces Thereover, whether over the canonical phase space or over a space of spacetimes.
The canonical case's zero-commutant conditions, moreover, can be readily converted, for the practicalities of solution, 
into a system of PDEs to which the {\bf Flow Method} applies. 
More specifically, we consider Lie's Integral Method of Invariants, and its slight modification to the Integral Method for Observables.  

\m 

\n Constraint Closure can moreover be extended by Feeding Families of Candidate Theories -- deformation of algebraic structures -- into a Dirac-type Algorithm.
This results in Rigidity returning details of GR's dynamics, and locally-SR spacetime structure thus amounting to Spacetime Construction \cite{RWR, AM13, ABook}.   
This can once again be conceptualized in Lie-theoretic terms, as {\bf Lie Algebraic Rigidity}, 
now embracing not only Quantum Spacetime Reconstruction but even the Foundations of Geometry \cite{A-Brackets}.    

\m

\n Refoliation Invariance can be expressed as a `commuting pentagon' Lie algebraic structure condition. 
The Dirac algebroid formed by GR's constraints furthermore satisfies these requirements.   

\m 

\n Among the above uses of Lie's Mathematics, moreover, Lie brackets algebraic structures see particularly major and central use. 
These are moreover now argued to run not on Dirac's Mathematics but by Lie's more general Mathematics, 
through examples of `Dirac magic' having been observed \cite{A-Brackets} to be realized by Lie brackets more generally than in Dirac's classical Poisson brackets setting.

\subsection{Frontiers}

\n Subsequent series of Articles on A Local Resolution of the Quantum Problem of Time 
                and on the classical Global Problem of Time are imminent.  
TRiCQT, TRiPIQT and semiclassical resolution are already out -- see Part III of \cite{ABook} -- 
as are some global parts of the Comparative Theory of Background Independence: \cite{ABook}'s Epilogues and \cite{A-Killing, A-Cpct, A-CBI}.

\section{Configurations and configuration spaces $\FrQ$}\label{Q-Primary}

\n{\bf Definition 1} {\it Configurations} \cite{Lanczos, Arnold}
\be 
\biQ  \m \mbox{ with components } \m  Q^A  \m ,
\ee 
 are instantaneous snapshots of the state of a system $\lFrs$.

\m 

\n{\bf Definition 2} The space of all possible configurations $\biQ$ for a given system $\lFrs$ is the corresponding 
{\it configuration space} \cite{Lanczos, Arnold, ABook},
\be 
\FrQ(\lFrs)  \m ;  
\ee
{\bf Notation 1} The dimension of configuration space is 
\be
k := \mbox{dim}(\FrQ(\lFrs))
\ee 
\n{\bf Notation 2} In the current Series, we use slanted font for finite-dimensional entities and straight font for field entities.  

\m 

\n{\bf Notation 3} We use mathfrak font for the corresponding spaces of entities. 
This is as a means of immediately avoiding confusion between objects of a given type and the spaces thereof. 

\m 

\n{\bf Remark 1} In Articles I to V, we use Finite Theory as a default, with full GR as the only Field-Theoretic exception. 
From Article VI onward, Electromagnetism, Yang--Mills Theory and full GR coupled to whichever of these or a scalar field are considered, 
so the portmanteau notation starts in earnest in that Article.  

\m 

\n{\bf Remark 2} This Sec's examples are moreover unreduced, to Sec II.6's further examples of configuration spaces being reduced.

\subsection{Example 1) Newtonian Mechanics}

\n{\bf Definition 3} For $N$ particles in the usual flat-space $\mathbb{R}^d$ model of absolute space, 
we denote the incipient configurations -- {\it N-point constellations} -- by  
\be 
\biq \m \mbox{ with components } \m q^{aI}  \m , 
\ee 
for $a$ a $\mathbb{R}^d$ vector index running from 1 to $d$ (such vectors are also denoted by underlining) 
                         and $I$ a particle               label running from 1 to $N$.  

\m

\n{\bf Definition 4} The corresponding configuration spaces are {\it constellation spaces} 
\be 
\FrQ(d, N)  \:=  \FrQ(\mathbb{R}^d, N) 
            \:=  \bigtimes_{I = 1}^N \mathbb{R}^d  \m . 
\ee 
\n{\bf Remark 1} Straightforwardly, 
\be 
\FrQ(\mathbb{R}^d, N)  \es  \{\mathbb{R}^{d}\}^N  
                       \es  \mathbb{R}^{d \, N}                 \m ,  
\ee
which is furthermore equipped with the standard Euclidean inner product or metric.

\subsection{Example 2) Full GR}

\n{\bf Structure 1} GR spacetime is 
\be 
\mbox{ a semi-Riemannian 4-metric } \m  \bg \m \mbox{ with components } \m  \mg_{\mu\nu}(\vec{X}) 
\ee 
on a
\be 
\mbox{ 4-$d$ topological manifold } \m \FrM                                                         \m ;
\ee  
we use the spacetime vector 
\be 
\vec{X} \m \mbox{ with components } \m X^{\mu} \m \mbox{ to denote spacetime coordinates}           \m .   
\ee 
\n{\bf Structure 2} Therein, the incipient configurations are 
\be 
\mbox{Riemannian 3-metrics } \m \bh \m  \mbox{ with components } \m \mh_{ab}(\u{x})  
\ee 
on a fixed 
\be 
\mbox{ 3-$d$ topological manifold } \m \bupSigma                                                    \m ;  
\ee
interpreted as a spatial slice of GR spacetime;  we use the space vector
\be 
\underline{x} \m \mbox{ with components } \m  x^a \m \mbox{ to denote spacetime coordinates}        \m .  
\ee 
\n We will refer to this point of view as GR-as-Geometrodynamics.  
It is the most longstanding dynamical \cite{Darmois, B52} and canonical \cite{ADM, WheelerGRT, Battelle, DeWitt67} formulation of GR.  

\m

\n{\bf Modelling assumptions} We consider $\bupSigma$ compact without boundary (CWB) and connected.  
3-spheres $\mathbb{S}^3$ and 3-tori $\mathbb{T}^3$ are the most commonly considered specific spatial topological manifolds in the geometrodynamical literature, 
of which we make later use of $\mathbb{S}^3$ specifically.  

\m 

\n{\bf Remark 1} As the 3-metric $\bh$ is a symmetric $3 \times 3$ matrix, it has 6 degrees of freedom per space point. 
 
\m  
																						  
\n{\bf Definition 1} The space formed by the totality of the $\mh_{ab}$ on a fixed $\bupSigma$ is GR's incipient configuration space 
\be 
\FrQ(\bupSigma) = \Riem(\bupSigma)                        \m .   
\ee
\n{\bf Remark 2} The 3-metrics $\mh$ are analogous to Newtonian Mechanics' incipient $N$-point configurations $\biq$, 
with $\Riem(\bupSigma)$ playing an analogous role to constellation space $\FrQ(d, N)$, 
and $\bupSigma$ playing a more loosely analogous role to the underlying absolute space $\mathbb{R}^d$, 
See Articles II and VI for more about GR configuration spaces.

\subsection{Example 3) Minisuperspace GR}

\n{\bf Structure 1} Homogeneous positive-definite 3-metrics are notions of space in which every point has the same properties.  

\m 

\n{\bf Structure 2} The space of all of these on a given spatial topology $\bupSigma$ is {\it Minisuperspace} \cite{mini, Magic}
\be 
\Mini(\bupSigma) \m ; 
\ee  
This is a simpler, particularly symmetric subcase of GR.  

\m

\n The specific Minisuperspace models used in this Series of Articles' detailed examples are spatially closed on Machian grounds.
I.e.\ these avoid undue influence of boundary or asymptotic physics, a criterion that Einstein also argued for \cite{Ein21}.  
%

\m 

\n{\bf Example 1} The simplest choice is $\bupSigma = \mathbb{S}^3$; this is also the most conventional for closed-universe cosmologies. 
%
%
One needs at least 2 degrees of freedom, and Cosmology conventionally makes use of scalar fields alongside the scalefactor of the universe, $a$. 
The simplest case brings in one minimally-coupled scalar field.  
The $\FrQ$ metric for this Minisuperspace is (up to a conformal factor of $a^3$) just 2-$d$ Minkowski spacetime $\mathbb{M}^2$ equipped with its standard indefinite flat metric.  

\m 

\n{\bf Example 2} Other models considered accompany the scalefactor $a$ with anisotropies: 2 $\beta_{\pm}$ 
                                                                        or the larger set of $\u{\u{\beta}}$ (of which there are 5, by tracelessness).

\m 

\n{\bf Structure 2} Let us also introduce
\be 
\Ani(\bupSigma)
\ee 																		
to denote the {\it anisotropyspace} that these form (this Series only considers the $\beta_{\pm}$ case in any detail).

\subsection{Point transformations}\label{Q-Geom}

For subsequent Articles' use, $\FrQ$'s morphisms -- the coordinate transformations of $\FrQ$ -- are termed the {\it point transformations} and form the mapping space 
\be 
Point(\FrQ)  \m .
\ee

\section{Interlude on the Standard Principles of Dynamics}

This and the next two Appendices support facets 1 to 4 of the Problem of Time.  
Consult \cite{Goldstein, Lanczos} as preliminary reading if unfamiliar with this material.

\subsection{The Euler--Lagrange action}

\n{\bf Structure 1} Given the configurations $\biQ$ indexed by $\fA$, various derived objects can be built up from these.
The first we require are {\it velocities}                 
\be 
{\biQ}^{\prime}  \:=  \frac{\d \biQ}{\d t}                       \m , 
\ee 
for $t$ the Newtonian time $t^{\sN\se\sw\st\so\sn}$ in Mechanics or a coordinate time $\mt^{\scc\so\so\sr\sd}$ in GR.   

\m 

\n{\bf Structure 2} {\it Lagrangian variables} \cite{Lagrange} are then 
\be 
(\biQ, \biQ^{\prime})                                                 \m .   
\ee
\n{\bf Structure 3} It is enlightening to furthermore view these derived objects as forming the tangent bundle over configuration space, 
\be 
\FrT(\FrQ)                                                            \m . 
\ee  
\n{\bf Structure 4} {\it Kinetic metrics} $\u{\u{\biM}}$ with components $M_{\sfA\sfB}(\biQ)$ are a further type of composite object which feature in the theory's kinetic term, 
\be 
T  \:=  \half ||\biQ^{\prime}||_{\sbiM}\mbox{}^2  
   \:=  \half M_{\sfA\sfB} Q^{\sfA \, \prime} Q^{\sfB \, \prime}      \m . 
\label{T-1}
\ee 
This can additionally be considered to equip the configuration space $\FrQ$ with a metric, which the current Series assumes to be time and velocity independent 
in its fundamental whole-universe setting.  

\m 

\n{\bf Modelling Assumption} This is for a finite second-order classical physical system \cite{Lanczos, Goldstein} expressed in Lagrangian variables.  

\m 

\n The kinetic term can furthermore be viewed as a mapping 
\be 
\FrT(\FrQ) \longrightarrow \mathbb{R}                                 \m . 
\label{TT-R}
\ee 
\n{\bf Structure 5} The {\it potential term} is 
\be 
V = V(\biQ)                                                           \m :  
\ee
also time and velocity-independent in the intended fundamental whole-universe setting. 
In the case of Mechanics, this is otherwise an a priori free function, but takes a more specific form in the case of GR.  

\m

\n{\bf Structure 6} All dynamical information is contained within the {\it Lagrangian} function $L(\biQ, \biQ^{\prime})$.  
The most common form this takes is 
\beq
L(\biQ, \biQ^{\prime})  \es  T(\biQ, \biQ^{\prime}) - V(\biQ) 
\eeq 
for $T$ as given by (\ref{T-1}).  
 
\m

\n{\bf Structure 7}  The familiar difference-type Euler--Lagrange action for a finite theory is    
\be 
{\cal S}_{\sE\sL}  \es  \int L(\biQ, \biQ^{\prime}) \,                             \d t
                   \es  \int \{T(\biQ, \biQ^{\prime}) - V(\biQ)\}                                \d t 
                   \es  \int \left\{ \half {||\d \biQ^{\prime}||_{\sbiM}}^2 - V(\biQ) \right\}   \d t  \m , 
\label{S-EL}
\ee 
The bundle map interpretation of this parallels (\ref{TT-R}).

\m 

\n{\bf Example 1} The most usual Mechanics version \cite{Euler} of this has position vector coordinates 
\be 
\biQ = \biq \m \mbox{ with components } \m q^{Ia}                                         \mma  
\ee
and diagonal constant-mass kinetic metric
\be
M_{IaJb} = m_I\delta_{IJ}\delta_{ab}                                                      \m .  
\ee

\subsection{Euler--Lagrange equations and some of their common simplifications}

Next apply the standard prescription of the Calculus of Variations to obtain the equations of motion such that ${\cal S}_{\sE\sL}$ 
is stationary with respect to the $\biQ$.
This approach considers the true motion between two particular fixed endpoints $e_1$ and $e_2$ 
alongside the set of varied paths about this motion (subject to the same fixed endpoints).  
It gives rise to the {\it Euler--Lagrange equations},  
\be
\frac{\d }{\d t} \left\{ \frac{\pa L}{\pa \biQ^{\prime}} \right\}  \es  \frac{\pa L}{\pa \biQ}  \m .
\label{ELE}
\ee
These equations simplify in the three special cases below, two of which involve particular types of coordinates.
Indeed, one major theme in the Principles of Dynamics is judiciously choosing a coordinate system with as many simplifying coordinates as possible.  

\m

\n{\bf Simplification 1} {\it Lagrange multiplier coordinates} $\bim$ are such that $L$ is independent of $\bim^{\prime}$, 
$$
\frac{\pa L}{\pa\bim^{\prime}}  \es  0  \m . 
$$
The corresponding Euler--Lagrange equations then simplify to  
\be
\frac{\pa L}{\pa \bim}  \es  0          \m .
\label{lmel}
\ee
{\bf Simplification 2)} {\it Cyclic coordinates} are such that $L$ is independent of $\bic$ itself, 
$$ 
\frac{\pa L}{\pa \bic}  \es  0          \m ,
$$ 
but features $\bic^{\prime}$: the corresponding {\it cyclic velocities}.
The $\bic$ Euler--Lagrange equations then simplify to 
\be
\frac{\pa L}{\pa \bic^{\prime}}  \es  \mbox{\bf const}                  \m .
\label{cyclic-vel}
\ee
{\bf Simplification 3)} {\it The energy integral type simplification}. 
If $L$ is free from the independent variable $t$,  
$$
\frac{\pa L}{\pa t}  \es  0                                             \m ,
$$
then one Euler--Lagrange equation may be supplanted by the first integral
\be
L - \biQ^{\prime} \cdot \frac{\pa L}{\pa \biQ^{\prime}}  \es  constant  \m .
\label{en-int}
\ee

\subsection{Multiplier elimination} 

Suppose that we can 
\be
\mbox{solve } \m  0  \es  \frac{\pa L}{\pa \bim}(\bar{\biQ}, \bar{\biQ}^{\prime}, \bim) \m \mbox{ as equations for the } \m \bim   \m .  
\label{LME}
\ee
\n{\bf Remark 1} These equations arise from simplification 1), and $\bar{\biQ}$ denotes the system's non-multiplier coordinates.

\m 

\n Solvability is not in general guaranteed.
Firstly, (\ref{LME}) can on occasion be not even well-determined due to some of the $\bim$ being absent from the equations  
                                                              or due to some equations not being independent. 

\m 
															  
\n Secondly, it is also possible for (\ref{LME}) to admit no solution (or only a non-real solution which cannot be applied physically, 
or a solution that is not in closed form \cite{FileR}). 

\m

\n In the absence of these pathologies,  
\be 
\mbox{\it multiplier elimination}: \m  L(\bar{\biQ}, \bar{\biQ}^{\prime}, \bim) \longrightarrow  L_{\sr\se\sd}(\bar{\biQ}, \bar{\biQ}^{\prime})  \m .   
\ee

\subsection{Conjugate momenta}

{\bf Structure 1} We now consider further derived objects: the {\it conjugate momenta},  
\be 
\biP  \:=  \frac{\pa L}{\pa \biQ^{\prime} }  \m . 
\label{mom-vel}
\ee
Explicit computation of this for (\ref{S-EL}) gives the {\it momentum--velocity relation}
\be
\u{\biP} = \u{\u{\biM}} \cdot \u{\biQ}^{\prime}                  \m .
\ee
N.B.\ that $\biM$ is a matrix, so one would need two dot products to get a scalar output.   
The definition of $\biP$ enables further formulation of the preceding Section's simplifications. 

\m 

\n {\bf Simplification 1} Now the preliminary condition in deducing the multiplier condition is 
\be 
\dot{\biP}^{\scc} = 0  \m . 
\ee 
\n {\bf Simplification 2} The cyclic coordinate condition is 
\be
\biP^{\scc} = \mbox{ constant }  \m .  
\label{cyclic-vel-P}
\ee
\n {\bf Simplification 3} The energy integral is 
\beq
L - \dot{\biQ} \cdot \biP = \mbox{ constant }  \m .
\eeq

\subsection{Legendre Transformations}\label{Legendre}

{\bf Definition 1} Suppose we have a function 
\be 
F(\biy, \biv)
\ee 
and we wish to use 
$$
\biz  \:=  \frac{\pa F}{\pa \biy}
$$
as variables in place of the $\biy$.
To avoid losing information in the process, a {\it Legendre transformation} is required: passing to a function  
\beq
G(\biz, \biv)  =  \biy \cdot \biz - F(\biy, \biv)  \m .
\eeq
\n{\bf Remark 1} Legendre transformations are symmetric between $\biy$ and $\biz$: if one defines  
$$
\biy  \:=  \frac{\pa G}{\pa \biz}  \m ,
$$ 
the reverse passage yields 
$$
F(\biy, \biv) = \biy \cdot \biz - G(\biz, \biv)  \m .
$$  
\n{\bf Example 2} Suppose that our function is a Lagrangian $L(\biQ, \biQ^{\prime})$ and that 
we wish to use some of the conjugate momenta $\biP$ as variables in place of the corresponding $\biQ^{\prime}$.

\subsection{Passage to the Routhian}\label{Routh}

{\bf Example 3} (of Legendre transformation). Start from a Lagrangian with cyclic coordinates 
\be 
\bic \mma  L(\bar{\biQ}, \bar{\biQ}^{\prime}, \bic^{\prime})  \m ,
\ee  
for $\bar{\biQ}$ now the non-cyclic coordinates. 

\m 
 
\n{\bf Step 1} Exchange the $\bic^{\prime}$ for the corresponding momenta using (\ref{cyclic-vel-P}). 
\n This entails being able to 
\be
\mbox{solve } \m \m  \mbox{\bf const}  \es  \bip^c  
                                       \es  \frac{\pa L}{\pa \bic^{\prime}}(\bar{\biQ}, \bar{\biQ}^{\prime}, \bic^{\prime}) \m \m \mbox{ as equations for the $\bic^{\prime}$ } . 
\label{Routhian-Reduction}
\ee
\n{\bf Step 2} Unlike in multiplier elimination, however, these are not just to be substituted back into the Lagrangian.
Rather, one additionally needs to apply the Legendre transformation (\ref{Routhian}), by which one passes not to a na\"{\i}ve reduced Lagrangian but to a {\it Routhian}   
\beq
R(\bar{\biQ}, \bar{\biQ}^{\prime}, \bip^c)  \:=  L(\bar{\biQ}, \bar{\biQ}^{\prime}, \bic) - \bip^c \cdot  \bic^{\prime}  \m .
\label{Routhian}
\eeq 
The overall process is known as 
\be 
\mbox{\it passage to the Routhian} \m :   L(\bar{\biQ}, \bar{\biQ}^{\prime}, \bic^{\prime})  \longrightarrow  R(\bar{\biQ}, \bar{\biQ}^{\prime}, \bip^c)  \m .  
\ee
\n{\bf Remark 1} This entails treating the cyclic coordinates as a separate package from the non-cyclic ones.  
 
\m  
 
\n{\bf Remark 2} This can be a useful trick \cite{Goldstein}, most usually in the context of simplifying the Euler--Lagrange equations. 
If this reduction can be performed, we can furthermore use cyclic momenta's constant status (\ref{cyclic-vel-P}) to free a dynamical problem from its cyclic coordinates.
I.e.\ this completes the corresponding part of the integration of equations of motion.

\subsection{Passage to the Hamiltonian}\label{Hamiltonian}

{\bf Example 3} (of Legendre transformation). 
Replace {\sl all} the velocities $\biQ^{\prime}$ by the corresponding momenta $\biP$, to form the {\it Hamiltonian} 
\beq
H(\biQ, \biP) \:=  \biP \cdot \biQ - L(\biQ, \biQ^{\prime}) \m 
\eeq
The variables 
\be 
(\biQ, \biP)
\ee 
are subsequently termed {\it Hamiltonian variables}. 
The issue of whether such a replacement is always possible is postponed to Sec \ref{Constraints}.  

\m  

\n{\bf Subexample} For the Lagrangian (\ref{S-EL}), 
\beq
H  \es  \half ||\biP||_{\sbiN}\mbox{}^2 + V(\biQ)  \m .
\eeq
This Series concentrates on the $t$-independent notion of Hamiltonian. 
The equations of motion are {\it Hamilton's equations},  
\be
\biQ^{\prime}  \es    \frac{\pa H}{\pa \biP}       \mma 
\biP^{\prime}  \es  - \frac{\pa H}{\pa \biQ}       \m . 
\ee
For the Lagrange multiplier coordinates, the first half of the corresponding Hamilton's equations collapse to just 
\beq
\frac{\pa H}{\pa \bim}  \es  0  \m .  
\eeq
On the other hand, in the $t$-independent case (\ref{en-int}) becomes $H = const$.

\m 

\n{\bf Remark 1} Most phase spaces considered in this Series take the form of cotangent space
\be 
\FrT^*(\FrQ) \m .  
\ee
\n{\bf Remark 2} Further motivations for the Hamiltonian formulation include admission of systematic treatment of constraints due to Dirac 
(\cite{Dirac, HTBook} and Sec \ref{Constraints}) and its greater closeness to Quantum Theory.

\section{Temporal Relationalism (aspect 0a): Leibnizian implementation}\label{TR-Intro}

\subsection{Leibnizian Time(lessness) Principle}

\n{\bf Leibnizian Time(lessness) Principle} There is no time at the primary level for the universe as a whole \cite{L, B94I, ABook}. 

\m   

\n The following two-part selection principles give a mathematically-sharp implementation \cite{BB82, FileR} at the level of Principles of Dynamics actions.  

\m 

\n{\bf Temporal Relationalism i)}  Include no extraneous times or extraneous time-like variables \cite{FileR}. 

\m 

\n{\bf Temporal Relationalism ii)} Include no label times either \cite{FileR}. 

\m 

\n The Euler--Lagrange action (\ref{S-EL}) however makes reference to the extraneous notion of time $t$, by which Temporal Relationalism i) fails.

\subsection{Jacobi actions are Manifestly Reparametrization Invariant}

{\bf Structure 1}  We can however replace (\ref{S-EL}) with {\it Jacobi's action principle} \cite{Jacobi}.
A first form for this is 
\be 
{\cal S}_{\sJ}  \es  2         \int \sqrt{W \, T_{\lambda}}              \d \lambda 
                \es  \sqrt{2}  \int \sqrt{E - V} ||\dot{\biQ}||_{\sbiM}  \d \lambda                         \m .   
\ee 
The notation for this is as follows. 
$\lambda$ is a `label-time' parameter and 
\be 
\dot{\m}  \:=  \frac{\d }{\d \lambda}                                                                        \m .  
\ee
\n{\bf Structure 2} A primary notion of velocity is here defined as the derivative with respect to $\lambda$:  
\beq
\mbox{velocity}            \:=         \frac{\d \mbox{(configuration variable)}}{\d (\mbox{label time})}  
                 \m \mbox{ i.e.\ } \m  \frac{\d \biQ}{\d\lambda}                                             \m .
\eeq 
\n{\bf Structure 3} The {\it parametrized Lagrangian variables} are  
\be 
(\biQ , \m \dot{\biQ})                                                                                       \m .
\ee 
The tangent bundle $\FrT(\FrQ)$ is here realized as {\it configuration--parametrized-velocity space}.

\m 
 
\n It is now this version of velocities which enter the otherwise-standard definition of kinetic term, 
\be 
T_{\lambda}  \:=  \half {||\dot{\biQ}||_{\sbiM}}^2  
             \es  \half M_{\sfA\sfB} \dot{Q}^{\sfA} \dot{Q}^{\sfB}                                           \m .  
\ee 
We assume in this Series that this takes the most physically standard form: homogeneous quadratic in the velocities; this assumption is removed in Sec \ref{JSS}.  

\m 

\n{\bf Structure 3} $E$ is the {\it total energy} of the system, and the {\it potential factor} 
\be 
W(\biQ) := E - V(\biQ)  \m . 
\ee 
\n{\bf Structure 4} The {\it Jacobi action} \cite{Lanczos} is of the product form 
\beq
{\cal S}^{\sM\sR\sI}_{\sJ}  \:=    \int \d\lambda \, L^{\sM\sR\sI}_{\sJ}  
                            \es  2 \int \d\lambda \sqrt{T_{\lambda}W}       \m .   
\label{J-action}
\eeq 
\n{\bf Remark 1} The Jacobi action principle clearly complies with Temporal Relationalism i). 

\m 

\n{\bf Remark 2} It moreover complies with Temporal Relationalism ii) as well. 
This is by {\bf Manifestly Reparametrization Invariance}: switching to a monotonically-related\footnote{This 
inequality ensures no zero factors enter in the form of $d \mu/\d \lambda$ terms.} 
label-time $\mu$ gives an equivalent action by cancellation of the label-time coordinate changes by Fig \ref{TR-Implems}a). 
This rests on the kinetic term $T_{\lambda}$ being homogeneous-quadratic in the velocities, so $\sqrt{T_{\lambda}}$ is linear therein.  
Thus interchanging the action's parameter for any other monotonically related label does not alter the physical content of the theory.  

\m 

\n{\bf Remark 3} More generally, Manifest Reparametrization Invariance 
requires an action to be homogeneous-linear in its label-time velocities (Fig \ref{TR-Implems}.a). 

\m 

\n{\bf Structure 3} The $\d/\d \lambda$ in our parametrized velocities can moreover be viewed \cite{Stewart} as the Lie derivative 
\be
\pounds_{\frac{\d}{\d \lambda}}
\label{Lie-dot}
\ee 
in a particular frame.
%
{\begin{figure}[!ht]
\centering
\includegraphics[width=0.85\textwidth]{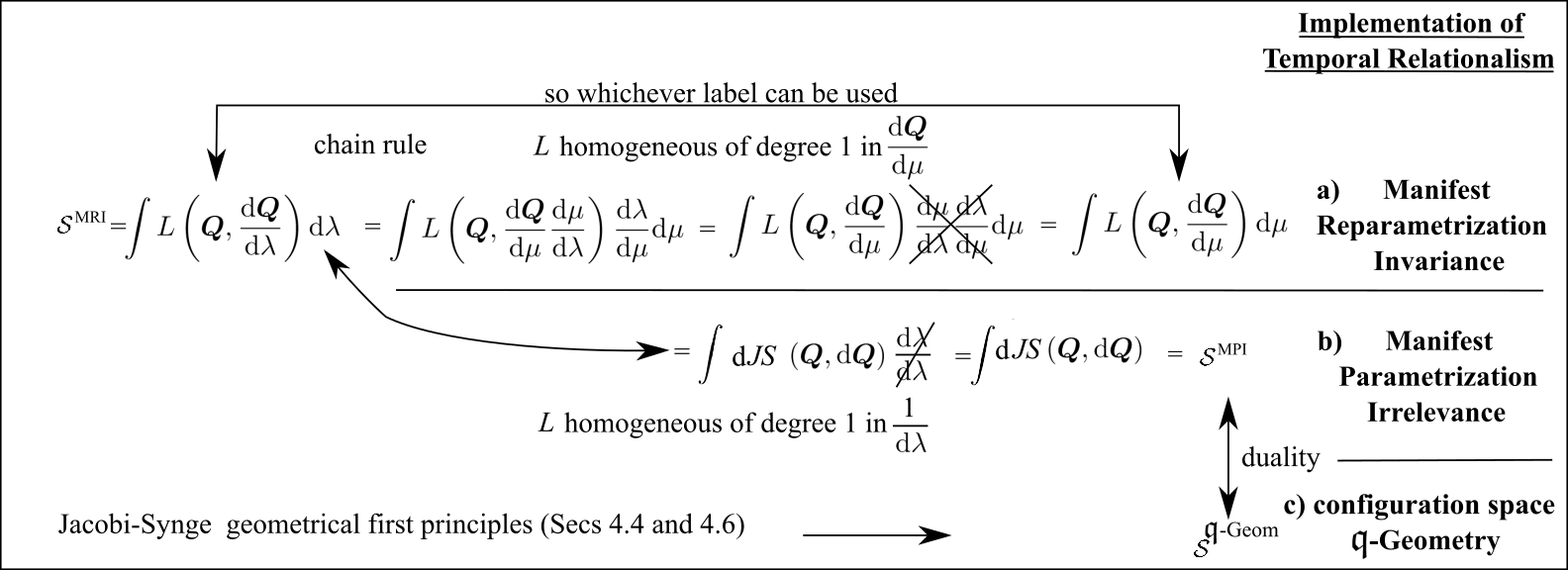}
\caption[Text der im Bilderverzeichnis auftaucht]{\footnotesize{Inter-relation of this Series of Articles' 
three implementations of Temporal Relationalism at the level of actions a-c).}} 
\label{TR-Implems}\end{figure}} 

\subsection{... or equivalently Manifestly Parametrization Irrelevant}

Here no use of $\lambda$ is to be made at all. 

\m 

\n{\bf Structure 1} One immediate consequence of this is that there is no primary notion of velocity: this has been supplanted by a {\it change in configuration} 
\be
\d \mbox{(configuration variable)} \m \mbox{ i.e.\ } \m \d \biQ     \m . 
\label{Change}
\ee 
{\bf Structure 2} {\it Configuration--change variables} are then 
\be 
(\biQ , \m \d\biQ)                                                  \m .  
\label{Q-dQ}
\ee 
These constitute an alternative representation of the tangent bundle $\FrT(\FrQ)$.

\m 

\n{\bf Structure 3} One is now to conceive in terms of {\it kinetic arc element} 
\be 
\d s := ||\d\biQ||_{\sbiM}  
     \es  \sqrt{ M_{AB} \d Q^A \d Q^B }
\label{hom-quad}
\ee 
in place of kinetic energy.  

\m 

\n{\bf Structure 4} One is additionally to conceive in terms of the {\it physical} alias {\it Jacobi arc element} 
\be 
\d J  \:=  \sqrt{ 2 W}      \,                \d s  
      \es  \sqrt{2} \sqrt{ E - V }  ||\d\biQ||_{\sbiM}
  	  \es  \sqrt{2} \sqrt{ E - V } \sqrt{M_{AB} \d Q^A \d Q^B}
\label{dJ}	   
\ee 
in place of Lagrangian.

\m 

\n{\bf Structure 5} On thus succeeds to formulate one's action without use of any meaningless label at all. 
This furnishes a second furtherly conceptually advanced implementation of Temporal Relationalism ii): the {\bf Manifestly Parametrization Irrelevant} form of the Jacobi action 
\be 
{\cal S}_{\sJ}^{\sM\sP\sI}  \es  \int \d J  
                            \es  \sqrt{2} \int \sqrt{W} \,                             \d s 
                            \es  \sqrt{2} \int \sqrt{W(\biQ)} ||\d \biQ||_{\sbiM} 
                            \es  \sqrt{2} \int \sqrt{E - V(\biQ)}  \sqrt{M_{AB}(\biQ)  \d Q^A \d Q^B}  \m .   
\label{GeneralAction}							
\ee  
Jacobi himself \cite{Jacobi} already formulated his action principle in this way.  

\m 

\n Actions are now more generally required to be homogeneous of degree one in the changes (this is clearly equivalent by Fig \ref{TR-Implems}.b).    

\m

\n{\bf Lemma 1}. (\ref{J-action}) and (\ref{GeneralAction}) are indeed equivalent. 

\m 

\n{\u{Proof}} The arrow between a) and b) of Fig \ref{TR-Implems}. $\Box$ 

\m 

\n{\bf Structure 6} The $\d$ in our changes can now be interpreted in terms of the Lie derivative \cite{Pauli, Sleb, Yano55, Yano70}
\be 
\pounds_{\d}
\ee 
in a particular frame (paralleling \ref{Lie-dot}).

\subsection{... or, dually, a geometrical action}

It is moreover a further conceptual advance for Background Independent Physics to cease to employ names or notions 
deriving from physically irrelevant or Background Dependent entities.  

\m 

\n In the present case, this involves ceasing to even mention any meaningless label or parameter. 
This can be done because the Manifestly Parametrization Irrelevant implementation is, dually, a {\bf (Configuration Space) Geometric} Implementation, 
\be 
{\cal S}_{\sJ}^{\sFrQ\mbox{-}\sG\se\so\sm}  \es  \sqrt{2} \int \sqrt{W(\biQ)} ||\d \biQ||_{\sbiM}  \m .   
\ee 
\n{\bf Remark 1} This view of our action is commonplace in the Dynamics \cite{Arnold} and Celestial Mechanics literatures in precisely this dual geometrical action conception.  
This does however obscure its Manifestly Parametrization Invariant meaning, which more directly addresses the relevance of such an action to the theory of time.  

\m 

\n{\bf Remark 2} The premise of a geometrical action itself is to view dynamics as a geodesic on the corresponding configuration space geometry. 

\m 

\n{\bf Remark 3} We use ${\cal S}_{\sJ}$ to mean either interpretation: ${\cal S}_{\sJ}^{\sFrQ\mbox{-}\sG\se\so\sm}$ or ${\cal S}_{\sJ}^{\sM\sP\sI}$.

\subsection{Euler--Lagrange to Jacobi action as a passage to the Routhian}

\n{\bf Remark 1} The passage from Euler--Lagrange's action principle with no explicit $t$ dependence to Jacobi's action principle 
(\ref{J-action}) is a subcase of passage to the Routhian.   

\m 

\n For rewrite $S_{\sE\sL}$ as the $\lambda$-parametrization of the action arising by appending time to the system's configurations: 
\be
{\cal S}_{\sE\sL}^{\sP\sa\sr\sa\sm} := \int \d \lambda \, \dot{t} \, L(\biQ, \dot{\biQ})   \m . 
\label{Ltadjac}
\ee
\n{\bf Remark 2} The original Lagrangian's explicit $t$-independence means that $t$ in (\ref{Ltadjac}) is a cyclic coordinate. 

\m 

\n Passage to the Routhian thus applies, yielding  
\be
L_{\sT\sR}(\biQ, \dot{\biQ})  \:=  L(\biQ, \dot{\biQ})\dot{t} - P^t \dot{t}  \m \m \mbox{  for } \m \m 
\frac{\pa L}{\pa \dot{t}}     \es   P^t 
                              \es   - E                                       \m , \m \mbox{ constant } .  
\label{tprimeeq}
\ee
\n{\bf Remark 3} See Sec \ref{JME} for the converse working. 

\m 

\n{\bf Remark 4} This working is termed the {\it parametrization procedure} in the Mechanics literature \cite{Lanczos}. 
This refers to the (nonrelational!) adjunction of the 1-$d$ space of a time variable to the configuration space 
\be 
\FrQ \longrightarrow \FrQ \times \mathbb{R}  \m .
\ee
\n{\bf Remark 5} This working also serves to justify the identification of the Jacobi action's potential factor $W$ as the combination of well-known physical entities $E - V$.

\subsection{Jacobi--Synge actions}\label{JSS}

{\bf Structure 1} More generally, Temporal Relationalism is implemented by 
\be 
{\cal S}_{\sJ\sS}  \es  \int L_{\sJ\sS} \d \lambda  
                   \es  \int \d JS
\ee
with Jacobi--Synge Lagrangian                     $L_{\sJ\sS} = L_{\sJ\sS}^{\sM\sR\sI}$  that is homogeneous-linear in $\dot{\biQ}$ 
and  Jacobi--Synge arc element \cite{Synge, Lanczos} $\d JS      =      \d JS^{\sM\sP\sI}$, 
                                       or dually, $\d JS^{\sFrQ\mbox{-}\sG\se\so\sm}$     that is homogeneous-linear in $\d{\biQ}$.  
The equivalences in Fig \ref{TR-Implems} are already stated at this level of generalization.\footnote{Synge's generalization 
of Jacobi parallels Finsler's (and in fact Riemann's) generalization of Riemannian Geometry, 
although Synge's considerations are furtherly general in not requiring nondegeneracy.}    

\m 

\n{\bf Structure 2} The {\it Jacobi--Synge construction} is, given a configuration space metric geometry, to form the corresponding theory of Mechanics thereupon.

\subsection{Hamiltonian formulation for constrained systems}\label{Constraints}

{\bf Structure 1} Passage from the Lagrangian to the Hamiltonian formulations can be nontrivial. 
For the {\it Legendre (transformation) matrix}  
\beq
\u{\u{\bslLambda}}  \:=  \frac{  \pa^2 L  }{  \pa \u{\biQ}^{\prime} \pa \u{\biQ}^{\prime}  }      \m 
\left( 
                               \es  \frac{  \pa \u{\biP}  }{  \pa {\u{\biQ}}^{\prime}  } 
\right)  
\label{Leg-Matrix}
\eeq
-- named by its latter form being associated with the Legendre transformation -- is in general non-invertible. 
Thereby, the momenta $\biP$ cannot be independent functions of the velocities $\biQ^{\prime}$.

\m 

\n{\bf Example 1} In the case of the action with purely-quadratic kinetic term, the Legendre matrix is just the kinetic matrix $\biM$. 
%

\m  

\n{\bf Definition 1} In the Hamiltonian formulation, {\it constraints} are relations 
\beq 
\sbcC(\biQ, \biP) = 0
\eeq 
between the momenta $\biP$ by which these are not independent.
These carry $\fC$ indices when required, using primed and double-primed versions when multiple concurrent such are required in a given formula.
This is our general notation for specialized-object indices: the letter says which type of specialized object, 
with priming used to cover multiple uses of such indices within a single formula.  
This is quite a general type of constraint considered by Dirac \cite{Dirac}.

\m 

\n{\bf Remark 1} The Euler--Lagrange equations can be rearranged to reveal the explicit presence of the Legendre matrix, 
\beq
\u{\u{\bslLambda}} \cdot \u{\biQ}^{\prime\prime}  \es  \frac{  \pa^2 L  }{  \pa \biQ^{\prime}  \pa \biQ^{\prime}  } \cdot \biQ^{\prime\prime}   
                                                  \es  \frac{  \pa L  }{  \pa \biQ  }  -  
												       \frac{  \pa^2 L  }{  \pa \biQ \pa \biQ^{\prime}  }  \cdot  \biQ^{\prime}                           \m .
\label{acc}
\eeq
The above noninvertibility additionally means that the accelerations are not uniquely determined by the Lagrangian data $(\biQ, \biQ^{\prime})$.

\m 

\n{\bf Notation 1} Keeping track of which objects enter a Background Independence scheme alias Problem of Time resolution is moreover an issue. 
For this requires many objects, many of which are new to address aspects or facets on the one hand, 
or newly formulated or interpreted because traditional versions of them succumbing to facet interference \cite{K92, I93} on the other.  
Because of this, we jointly denote constraints -- a sizeable subclass of such objects -- by the undersized calligraphic font, 
so that constraint status can immediately be read off the formalism.  

\m 

\n{\bf Definition 2} Constraints arising from the above non-invertibility of the momentum--velocity relations alone are termed  {\it primary} \cite{Dirac, HTBook}, denoted by 
\be 
\bscP \mma \mbox{ indexed by } \m \fP   \m . 
\ee     
\n{\bf Definition 3} Constraints furthermore requiring input from the variational equations of motion are termed {\it secondary} \cite{Dirac, HTBook}, denoted by 
\be 
\bscS \mma \mbox{ indexed by } \m \fS   \m . 
\ee 
[Classification of constraints into primary and secondary is originally due to Bergmann; it is not reformulation independent \cite{HTBook}.]

\m 

\n{\bf Remark 2} Constraints arising from the propagation of existing constraints using the equations of motion are an intuitively valuable case of secondary constraints 
(albeit these are on some occasions called `tertiary constraints').
%

\m 

\n{\bf Remark 3} We often use the parametrized-velocity, alias dotted, version of this subsection, i.e.\ using $\dot{\biQ}$ and $L_{\sJ\sS}$.

\subsection{Primary constraints from Temporal Relationalism}\label{Prim-TR}

We work for now in the Manifestly Reparametrization Invariant case. 

\m 

\n{\bf Definition 1} In this implementation, the definition of momentum takes the form 
\be 
\biP  \:=  \frac{\pa L_{\sJ\sS}}{\pa \dot{\biQ}}  \m .   
\label{P-MRI-Def}
\ee
{\bf Lemma 2 (Dirac)} \cite{Dirac} Manifestly Reparametrization Invariant actions imply at least one primary constraint.

\m 

\n{\u{Proof}} These are homogeneous of degree 1 in the velocities $\dot{\biQ}$.  

\m 

\n The $k := \mbox{dim}(\FrQ)$ conjugate momenta $\biP$ are consequently (\ref{P-MRI-Def}) homogeneous of degree 0 in $\dot{\biQ}$. 

\m 

\n I.e. functions of at most $k - 1$ ratios of velocities. 

\m 

\n There must thus be at least one relation between the momenta themselves, without any use having been made of the equations of motion.  

\m 

\n But this meets the definition of primary constraint.  $\Box$

\m 

\n{\bf Remark 1} Temporal Relationalism thus acts as a {\it Constraint Provider}: an underlying principle that produces constraints \cite{Battelle, B94I}.\f{This is 
in opposition to the `Applied Mathematics' point of view that constraints just are, no questions asked. 
Which opposition is made, in particular, in the context of investigating origins for {\sl fundamental theories' constraints}. 
This is furthermore an example of Wheeler asking for `zeroth principles' \cite{Battelle} whenever presented with `first principles'.} 
 
\m 

\n{\bf Example 1} In the particular case of Jacobi's action \cite{BB82, B94I}, the definition of momentum takes the form 
\be 
\u{\biP}  \:=  \frac{\pa L_{\sJ}}{\pa \dot{\biQ}} 
          \es  \sqrt{\frac{W}{T}} \u{\u{\biM}} \cdot \dot{\u{\biQ}}   \m . 
\ee
{\bf Lemma 3} (Barbour \cite{B94I}) For Jacobi-type actions, there is precisely one primary constraint, 
\beq
\scE  \es  \half ||\biP||_{\sbiN}\mbox{}^2 + V(\biQ) 
      \es                     E                       \m ,   
\label{E-Constraint}
\eeq
or, in terms of coordinates,
\be 
\scE  \:=  \half \, N^{AB} P_A P_B + V(\biQ)  
      \es                        E                    \m . 
\label{E}
\ee
Here 
\be 
\biN := \biM^{-1}  \m . 
\ee 
\n{\u{Proof}} 
\be  
||\biP||_{\sbiN}  \es  \left|\left| \sqrt{\frac{W}{T}} \biM \cdot \dot{\biQ} \right|\right|_{\sbiN} 
                  \es  \sqrt{\frac{W}{T}} || \dot{\biQ} ||_{\sbiM\sbiN\sbiM} 
                  \es  \sqrt{\frac{W}{T}} || \dot{\biQ} ||_{\sbiM}   
                  \es  \sqrt{\frac{W}{T}} \sqrt{2 \, T}
				  \es  \sqrt{2 \, W}                                                                 \m . \m \Box
\ee 
\n{\bf Remark 2} This can be envisaged as a `Pythagorean' or `direction-cosines' working. 
By this, $L_{\sJ}$'s quadraticness in its velocities induces  $\scE$'s    quadraticness in its momenta.

\m 

\n{\bf Remark 3} In the most common case of Mechanics (Temporally Relational but Spatially Absolute!), 
\be 
\scE  \es  \half ||\bip||_{\sbin}\mbox{}^2 + V(\biq) 
      \es                    E                                                      \m , 
\ee  
where the 
\be 
\mbox{{\it inverse constant-mass matrix} \m $\bin$ \m has components } \m  n_{IaJb} = \frac{1}{m_I}\delta_{IJ}\delta_{ab}                                    \m .    
\ee 
\n{\bf Remark 4} (\ref{E}) is much more common in the literature in a context in which it is appropriate to consider it a `constant-energy equation'. 
It however has a distinct interpretation in the current Temporally Relational whole-universe context, as per Sec \ref{EoT}.  

\m 

\n{\bf Remark 5} Taking into account this constraint causes one to pass from considering a point and a vector in $\FrQ$ to considering just a point and a direction.  
Thus one has not $\FrT(\FrQ)$ but a {\it direction bundle} alias {\it unit tangent bundle}, 
\be
\FrU(\FrQ) \m .  
\ee
Finally, in the current case, instead of the entities in question squaring to 1 as direction cosines do, 
the momenta `square' by use of the $\biM$ matrix's inner product to the `square of the hypotenuse', $2 \, W$.

\subsection{Equations of motion}\label{Evol-Eq}

These are 
\beq
\sqrt{\frac{W}{T}} 
\left\{
\sqrt{\frac{W}{T}} Q^{\sfA \, \prime}
\right\}^{\prime} 
+ \slGamma^{\sfA}\mbox{}_{\sfB\sfC} \sqrt{\frac{W}{T}} Q^{\sfB \, \prime}  
                                    \sqrt{\frac{W}{T}} Q^{\sfC \, \prime}   \es  
N^{\sfA\sfB}\frac{\pa W}{\pa Q^{\sfB}} \mbox{ } .
\label{MRI-ELE}
\eeq
where $\slGamma^{\sfA}\mbox{}_{\sfB\sfC}$ are the Christoffel symbols of the $\FrQ$ geometry.  

\m 

\n Or, in terms of momenta (\ref{MRI-ELE}) becomes 
\be
\sqrt{  \frac{W}{T} } \, {P}^{\sfA \, \prime} + \slGamma^{\sfA}\mbox{}_{\sfB\sfC} P^{\sfB} P^{\sfC}  \es  -  N^{\sfA\sfB}  \frac{\pa V}{\pa Q^{\sfB}}     \m .
\ee

\subsection{Outline of Manifestly Parametrization Irrelevant counterpart}

\n{\bf Remark 1} The Author \cite{FileR, ABook} showed moreover that all of the preceding subsection's arguments transcend to 
Manifestly Parametrization Irrelevant actions and to geometric actions dual thereto. 
Most details of this are postponed to Article V; the details we presently need are as follows. 

\m 

\n{\bf Structure 1} \n The notion of momentum $\biP$ carries over to this context. 

\m 

\n{\bf Remark 2} The formula defining generalized momentum does not, however, 
since (\ref{P-MRI-Def}) includes two parameter-bearing objects -- $\biQ^{\prime}$ and $L$.  

\m 

\n{\bf Definition 1} The Manifestly Parametrization Irrelevant formula for momentum is 
\be 
\biP  \:=  \frac{\pa \, \d JS}{\pa \, \d \biQ}   \m , 
\label{TRi-Mom}
\ee
i.e.\ the partial derivative of the Jacobi--Synge arc element with respect to the change.

\m 

\n{\bf Lemma 4} This formula is equivalent to the standard one. 

\m 

\n{\u{Proof}} The standard formula's parameter-bearing objects' parameters wash each other out, by the elementary `cancellation of the dots Lemma' \cite{Goldstein}. $\Box$

\m 

\n{\bf Example 1} Computing this out, in the case of the Jacobi action,  
\be
\biP  \es  \biM \frac{\sqrt{2 \, W} \, \d \biQ}{  || \d \biQ||_{\sbiM}  } \m .
\label{New-P-Compute}
\ee

\subsection{Model arena examples of Manifest Parametrization Irrelevance}

\n{\bf Example 1} {\it Jacobi's action principle} for Spatially-Absolute Mechanics has 
\be 
\d s_{\sJ}  =  \sqrt{ m_I\d q^I\d q^I } 
\ee
and 
\be 
W(\biQ) = E - V(\bq)   \m  ; 
\ee 
this is the example used in the current Section.

\m 

\n{\bf Example 2} {\it Misner's action principle} \cite{Magic} for Minisuperspace GR is also of this form.  
The Misner action is indeed a subcase of full Geometrodynamics' Baierlein--Sharp--Wheeler \cite{BSW} action (Sec II.3.1). 
Various subcases of the Misner action are considered in Sec \ref{MSS-Intro}.

\section{Temporal Relationalism: resolution by Mach's Time Principle}\label{MTP}

\n It is quite natural to ask whether there is a paradox between the Leibnizian Time(lessness) Principle, 
on the one hand, and, on the other hand, 
our appearing to `experience time', and this moreover featuring in many Laws of Physics that appear to apply in the Universe.   

\m 

\n{\bf Remark 1} Let us start by pointing to discrepancies between the two contexts. 

\m 

\n{\bf Discrepancy 1} Everyday experience concerns subsystems rather than the whole-Universe setting of this Leibnizian Principle.  

\m 

\n{\bf Discrepancy 2} Whereas `time' is a useful concept for everyday experience, the nature of `time' itself is in general less clear.

\m 

\n{\bf Mach's Time Principle} is that \cite{M} {\it ``It is utterly beyond our power to measure the changes of things by time. 
Quite the contrary, time is an abstraction at which we arrive through the changes of things."} 
I.e.\ `{\sl time is to be abstracted from change}'.  

\m 

\n{\bf Remark 2} Indeed, it is change that we directly experience, and temporal notions are merely an abstraction from that, albeit a very practically useful abstraction if carefully chosen.  

\m 

\n{\bf Remark 3} Mach's Time Principle thus {\sl resolves} the Leibnizian Time(lessness) Principle's timelessness at the primary level, by a secondary notion of time, i.e. an {\it emergent time}.  

\m 

\n{\bf Aside 1} See in particular \cite{K92, I93, ABook} for discussion of alternative strategies for handling Temporal Relationalism 
such as primary-level time (e.g.\ hidden or from appended matter), 
adhering to timelessness 
or supplanting time by a notion of history.

\subsection{Implementation i) Jacobi--Mach variables}

\n To implement Mach's Time Principle, we firstly, work in configuration--change variables $(\biQ, \d\biQ)$.
Indeed, combining the above Machian connotations with Jacobi's actual formulation of his action principle in terms of these, 
we now can justify using the further alias {\it Jacobi--Mach variables} for these.

\subsection{Implementation ii) Jacobi--Mach formula for momentum}\label{JMM}

\n{\bf Structure 1} The momentum--velocity relations are to be supplanted by {\it momentum--change relations}
\be 
\biP   =   \biP(\biQ, \d\biQ) 
      \es  \frac{\pa \, \d JS}{\pa \, \d \biQ} (\biQ, \d \biQ)  \m .  
\label{P-MPI-Def} 
\ee 
{\bf Example 1} For the Jacobi action itself,  
\be 
\u{\biP}  \:=  \frac{ \pa \, \d J }{ \pa  \, \d \u{\biQ} }  
          \es  \frac{ \sqrt{2 \, W} }{ \d s } \u{\u{\biM}} \cdot \d \u{\biQ}        \m . 
\ee 
{\bf Lemma 5} \cite{FileR} Manifestly Parametrization Irrelevant actions imply at least one primary constraint.

\m 

\n{\u{Proof}} These are homogeneous of degree 1 in the changes $\d \biQ$.  

\m 

\n The $k$ conjugate momenta $\biP$ are consequently (\ref{P-MPI-Def}) homogeneous of degree 0 in $\d \biQ$. 

\m 

\n I.e.\ functions of at most $k - 1$ ratios of changes. 

\m 

\n There must thus be at least one relation between the momenta themselves (i.e.\ without using the equations of motion).  

\m 

\n But this meets the definition of primary constraint.  $\Box$

\m 

\n Lemma 3 moreover admits the following Manifestly Parametrization Irrelevant {\u{Proof}}:   
\be  
||\biP||_{\sbiN}  \es  \left|\left| \frac{\sqrt{2 \, W}}{\d s} \biM \cdot \d \biQ \right|\right|_{\sbiN} 
                  \es  \frac{\sqrt{2 \, W}}{\d s} || \d \biQ ||_{\sbiM\sbiN\sbiM} 
                  \es  \frac{\sqrt{2 \, W}}{\d s} || \d \biQ ||_{\sbiM}   
                  \es  \frac{\sqrt{2 \, W}}{\d s} \d s 
				  \es  \sqrt{2 \, W}                                                                         \m . \m \Box
\ee

\subsection{Implementation iii) Equation of Time interpretation}\label{EoT}

\n{\bf Remark 1} Working with Temporal Relationalism implementing formulations is but the larger of two parts in handling Temporal Relationalism.  
Approaches using this eventually need to be completed by a Machian `time is to be abstracted from change' step, as follows.   

\m 

\n{\bf Interpretation 1} In the current context of being provided by Temporal Relationalism, the interpretation that $\scE$ is to receive is that of an {\it equation of time}.  

\m 

\n{\bf Interpretation 2} In particular, $\scE$ can be rearranged to give the {\it Jacobi emergent time},\f{In this Series of Articles, 
given a time variable $t$, we denote the `calendar year zero' adjusted version of this by the corresponding oversized $\lt := t - t(0)$.}  
by integrating 
\be 
\frac{\pa}{\pa t^{\se\sm(\sJ)}} 
      \:=  \sqrt{ \frac{W}{T} } \frac{\pa}{\pa\lambda}   \m ,  
\label{Ast-0}
\ee
to obtain 
\beq
\lt^{\se\sm(\sJ)}  \es  \int \d\lambda \sqrt{ \frac{T}{W} }  
                   \es  \int \frac{\d s}{\sqrt{2 \, W}}  
				   \es  \int \frac{||\d \biQ||_{\sbiM}}{\sqrt{2 \, W}}  \m . 
\label{t-em-J}
\eeq
The third form therein -- the Manifestly Parametrization Irrelevant or dual $\FrQ$-geometric formulation  -- 
is moreover manifestly an equation for obtaining time from change in direct compliance with Mach's Time Principle. 
More precisely, it is a formula for an emergent timefunction as an explicit {\sl functional} of change, schematically  
\be 
t^{\se\sm(\sJ)} = {\cal F}[\biQ, \d \biQ] \m .
\ee   
This entails interpreting the quadratic constraint not as an energy constraint in the usual sense but as an equation of time \cite{B94I, ARel2, FileR}. 
`em' stands for (classical) {\it emergent Machian} time, and `$J$' for Jacobi.  
Following Mach, this is ab initio a highly dependent variable rather than the independent variable that time is usually taken to be.
This is because this `usual' situation assumes that one knows beforehand what the notion of time to use, 
whereas the current position involves operationally establishing that notion (see the Conclusion for discussion).  

\m 

\n To celebrate this, let us term the type of constraint provided by Temporal Relationism a {\it Chronos constraint}, denoting all examples of such by 
\be 
\Chronos  =  0  \m . 
\ee

\subsection{Implementation iv) Jacobi--Mach equations of motion}\label{JME}

(\ref{MRI-ELE}) also make reference to times, velocities and Lagrangians.
In Manifestly Parametrization Irrelevant form, the equations of motion are, rather, the {\it Jacobi--Mach equations} that follow from Jacobi's arc element in terms of Machian variables:
\beq
\d \left\{ \frac{\pa \,\d J}{\pa \,\d \biQ}  \right\}  \es  \frac{\pa \, \d J}{\pa \biQ} \m \Rightarrow 
\label{JME-1}
\eeq
\beq
\frac{\sqrt{2 \, W}\, \d}{||\d \biQ||_{\sbiM}}  
\left\{
\frac{\sqrt{2 \, W} \, \d Q^{\sfA}}{||\d \biQ||_{\sbiM}}  
\right\} 
+ \slGamma^{\sfA}\mbox{}_{\sB\sfC} \frac{\sqrt{2 \, W} \, \d Q^{\sfB}}{||\d  \biQ||_{\sbiM}}  
                                 \frac{\sqrt{2 \, W} \, \d Q^{\sfC}}{||\d  \biQ||_{\sbiM}}    \es 
N^{\sfA\sfB}\frac{\pa W}{\pa Q^{\sfB}}                                                                             \m .
\label{New-Evol}
\eeq
\n{\bf Remark 1} (\ref{JME-1}) is an `impulse formulation' of Newton's Second Law.  

\m 

\n{\bf Remark 2} A final move -- useful in practical calculations -- 
involves supplanting one of the evolution equations by the emergent Lagrangian form of the quadratic constraint, 
\beq
\half M_{\sfA\sfB}\Last{Q}^{\sfA}\Last{Q}^{\sfB} + W  =  0                                                        \m .  
\label{ENERGY}
\eeq

\subsection{Discussion} 

{\bf Interpretation 1} (\ref{t-em-J}) moreover also implements the further principle \cite{MTW} of `choose time so that motion is simplest'.  
For, via
\beq 
\Ast  \:=  \frac{\pa}{\pa t^{\se\sm(\sJ)}} 
      \:=  \sqrt{ \frac{W}{T} } \frac{\pa}{\pa\lambda}   \m ,  
\label{Ast}
\eeq
it is also distinguished by its simplification of the momentum-velocity relations and equations of motion from (\ref{P-MPI-Def}) and (\ref{New-Evol}).
`abs' here denotes the standard differential-geometric absolute derivative.  
\beq
P_{\sfA}  =  M_{\sfA\sfB}\Last Q^{\sfB} \m , 
\eeq
\beq
D_{\sa\sb\sss}\mbox{}^2 {Q}^{\sfA}  \es  \Last\Last  Q^{\sfA} + \slGamma^{\sfA}\mbox{}_{\sfB\sfC}\Last Q^{\sfB}\Last Q^{\sfC} 
                                    \es  - N^{\sfA\sfB} \frac{ \pa V }{ \pa Q^{\sfA} }                                            \m . 
\label{parag}  
\eeq 
The latter is a {\it parageodesic equation} with respect to the kinetic metric (meaning it has a forcing term arising from the conformal $W$-factor). 

\m

\n{\bf Remark 1} In this way, Temporally-Relational Mechanics' emergent classical Machian time can be seen to amount to be a recovery of Newtonian time, 
but now on a Temporally-Relational footing. 

\m 

\n{\bf Remark 2} We furthermore split $\biQ$ into heavy slow $\bih$ and light fast $\bil$ parts, leading to an expansion of $t^{\se\sm}$ with 
$\bih$ part as leading term. 
This is a significant move to make as regards making contact with cosmological modelling.  

\m 

\n{\bf Remark 3} Article V has a stronger form of Remark 1 -- for (Temporally {\sl and Configurationally}) Relational Mechanics -- 
as well as a more detailed treatment of Remark 2's topic.    

\m 

\n{\bf Remark 4} One can finally posit the Euler--Lagrange principle, in terms of  $t^{\se\sm}$ as an action principle encoding these simplest-form equations.  

\m 

\n{\bf Remark 5} The arguments of Leibniz and Mach are philosophically compelling enough to apply to not just Mechanics but to Physics as a whole.

\section{Minisuperspace example}\label{MSS-Intro}

\subsection{Overview of the general minisuperspace model}\label{MSS-Overview}

The standard ADM-type minisuperspace action including a minimally-coupled scalar field is 
\be 
{\cal S}  \es  \int \d t \sqrt{h} \, \alpha \left\{ \frac{T_{\sG\sR}^{\sM\sS\sS}}{4 \, \alpha^2} + R  - 2\slLambda \right\}  \m , 
\ee  
is of Euler--Lagrange type. 
Here, $\alpha$ is the lapse, $h$ is the determinant of the spatial homogeneous metric $h_{sb}$ and $\slLambda$ is the cosmological constant, and we have, 
schematically 
\be 
T_{\sG\sR}^{\sM\sS\sS} = M_{AB}Q^A Q^B                                                                 \m .  
\ee 
\n{\bf Remark 1} One advantage of minisuperspace models over Mechanics models is that minisuperspace is a restriction of GR, 
so it inherits some features that Mechanics does not possess.

\m 

\n{\bf GR-like feature 1)} The kinetic metric $\biM$ is now indefinite.

\m 

\n{\bf GR-like feature 2)} More specific restrictions are imposed on the form of the potential than in RPMs.  

\m 

\n{\bf Remark 2} Minisuperspace modelling is additionally significant for the universe as a whole. 
Here, GR is more accurate than Mechanics, with even isotropic Minisuperspace constituting a highly accurate model for cosmological purposes.
%

\m 

\n See Articles II and III for some ways in which, conversely, suitably-relational Mechanics models have a distinct set of advantages over minisuperspace models.

\m 

\n The corresponding Jacobi-type action for minisuperspace is the {\it Misner-type action} \cite{Magic}, 
\be 
{\cal S}  \es  \half \int \d \lambda \, \sqrt{\overline{T} \, \overline{W}}    \m   
\ee
(here we use overlines to denote densitization: $\overline{O} = \sqrt{h} O$). 
Or, in geometrical form, 
\be 
{\cal S}  \es  \half \int \d s \, \sqrt{\overline{W}}                          \m ,  
\ee
for minisuperspace kintic arc element
\be 
\d s = ||\d \biQ||_{\sbiM}                                                \m ,
\ee  
\be 
W(\biQ) = R - 2 \, \slLambda
\ee 
(in the undensitized presentation).

\subsection{Isotropic model with scalar field}\label{Isotropic}

Here the action picks up $T_{\phi}$ and $V(\phi)$ pieces.
Specialization to closed $\mathbb{S}^3$ isotropic model with single scalar field has, 
\beq
\overline{T}  \:=  \mbox{exp}(3 \, \slOmega)  
\left\{
- \dot{\slOmega}^2 + \dot{\phi}^2
\right\}                                                                                                               \mma
\overline{W}  \:=  \mbox{exp}(3 \, \slOmega)\{\mbox{exp}(-2\slOmega) - V(\phi) - 2 \, \slLambda\}                                                                       \m . 
\label{MSS-Action}
\eeq
Here, the {\it Misner variable} 
\beq
\slOmega := \mbox{ln} \, a                                                                                             \m , 
\label{Misner} 
\eeq 
for $a$ the usual cosmological scale factor.  
The cosmological constant term $\slLambda$ therein is needed 
to support \cite{Rindler} the spatially-$\mathbb{S}^3$ FLRW cosmology with scalar field matter in the case in which matter effects are presumed small.  

\m  

\n{\bf Remark 3} On the one hand, from the ADM-type action, varying with respect to the lapse gives the minisuperspace version of the Hamiltonian constraint $\scH$. 

\m

\n On the other hand, from the Misner-type action, $\scH$ follows as a primary constraint, which is now conceptually a subcase of $\Chronos$.  

\m

\n The restriction to Minisuperspace of the GR Hamiltonian constraint $\scH$ now arises as a primary constraint,    
in an `indefinite triangle' version of Sec \ref{Prim-TR}'s `Pythagorean' or `direction-cosines' working.   
[Readers will already be familiar with `indefinite triangles' from studying SR or the hyperbolic functions.]

\m 

\n As we shall see in Article II, both of these manoeuvres carry over to full GR.  

\m 

\n{\bf Remark 4} The constraints $\scH$ and $\scE$ thus both illustrate that the distinction between primary and secondary constraints is artificial,
in so far as that the two are interchangeable under reformulation.  

\m 

\n{\bf Structure 2} The specific form taken by the Hamiltonian constraint is 
\beq
\scH  \:=  \half \, \mbox{exp}(-3 \, \slOmega) \big\{  - p_{\slOmega}^2 + p_{\phi}^2 + \mbox{exp}(6 \, \slOmega)\{V(\phi) + 2 \, \slLambda - \mbox{exp}(-2\slOmega)\} \big\}  
      \es  0                                                                                                                                                                     \m . 
\label{MSS-H}
\eeq
\n{\bf Structure 3} This can be rearranged to give the classical Machian emergent time,    
\beq
\lt^{\se\sm}  \es  \int  \sqrt{ \frac{- \d\slOmega^2 + \d\phi^2}{\mbox{exp}(-2\slOmega) - V(\phi) - 2 \, \slLambda} }  \m . 
\label{Mini-tem}
\eeq
\n{\bf Modelling Assumption 1)} The matter physics is light and fast ($l$) as compared to the gravitational physics being heavy and slow ($h$). 

\m 

\n{\bf Modelling Assumption 2)} More conventionally, the scalefactor and the homogeneous matter mode are jointly taken to be $h$, 
with only the below aniotropy or Article XI's inhomogenity playing subsequent $l$ roles.

\subsection{Anisotropic vacuum model}\label{MSS-Aniso}

{\bf Structure 1} The vacuum anisotropic cases \cite{mini, Magic, Ryan} whose configuration space metric is   
\beq
\d s^2 = - \d \slOmega^2 + \d\beta_+^2 + \d\beta_-^2 \m  .
\label{Bianchi-A}
\eeq
These models are potentially of great importance through being conjectured to be GR's generic behaviour near cosmological singularities \cite{BKL}.  

\m 

\n{\bf Structure 2} The potential term has the following specific form inherited from the densitized GR Ricci scalar potential term, 
\beq
\overline{V} = \mbox{exp}(\slOmega)\{V(\bbeta) - 1\} \mbox{ } , \mbox{ } \mbox{ for}
\label{B-IX-1}
\eeq
\beq
V(\bbeta_{\pm}) = \frac{\mbox{ exp}(-8 \, \beta_+)}{3} - \frac{4 \, \mbox{\scriptsize exp}(-2 \, \beta_+)}{3}  \mbox{cosh}\,(2\sqrt{3} \, \beta_-) + 1 
                + \frac{2\,\mbox{exp}(4 \, \beta_+)}{3}\{\mbox{cosh}(4\sqrt{3} \, \beta_-) - 1\}                                \m : 
\label{B-IX-2}
\eeq
an open-ended well of equilateral triangular cross-section. 

\m 

\n{\bf Structure 3} The Hamiltonian constraint is now 
\beq
\scH  \es  - p_{\Omega}^2 \, + \, p_+^2 \, + \, p_-^2 \, + \, \mbox{exp}(4 \, \slOmega)\{V(\beta_{\pm}) - 1\}  \m .  
\eeq
\n{\bf Structure 4} This can be rearranged to give the classical Machian emergent time,    
\beq 
\lt^{\se\sm(\sJ)}_{\mbox{\scriptsize isotropic-MSS}}  \es  \int \frac{ \mbox{exp}(\slOmega) \sqrt{- \d\slOmega^2 + \d\beta_-^2 + \d\beta_+^2}  }
                                                                     {  \sqrt{ 1 - V(\beta_{\pm}) }  }                                              \m . 
\label{Bianchi-IX-tem} 
\eeq
\n{\bf Modelling Assumption 3)} The anisotropy physics is light and fast ($l$) as compared to the scalefactor physics being heavy and slow ($h$). 

\m 

\n{\bf Remark 1} This and the previous subsection's models can readily be combined, including viewing slight anisotropy as a toy model for slight inhomogeneity.

\m 

\n{\bf Remark 2} Classical emergent Machian time from GR's Hamiltonian constraint amounts to a relational recovery of GR's version of proper time.  

\m 
 
\n In cases dominated by scalefactor (and other homogenous modes') dynamics, moreover, 
classical emergent time amounts to a relational recovery of cosmic time to leading order.

\section{Conclusion}

\n 1) We have considered a first aspect of Background Independence, namely {\it Temporal Relationalism}: 
the Leibnizian `there is no time for the universe as a whole at the primary level'. 
This is implemented into Physical Theory 

\m 

\n i) by involving neither extraneous time -- such as Newton's -- 
                                                                   nor extraneous time-like variable, e.g.\ GR's lapse $\upalpha$.

\m 

\n ii) By additionally not involving any label-time parameters either. 

\m 

\n Manifestly Reparametrization Invariant actions, such as Jacobi's for Mechanics or Misner's for minisuperspace GR, implement i) implicitly. 
Jacobi--Synge actions are totally general finite-theory such: homogeneous-linear in parameter-velocities $\d \biQ / \d \lambda$, 
with Jacobi and Misner's cases attaining this via a 'square root of a homogeneous-quadratic kinetic term' factor. 
Full (or indeed restricted but inhomogeneous) GR parallels this, but requires more work, 
with the Baierlein--Sharp--Wheeler (BSW) doing a partial job but Problem of Time facet interferences requiring the further relational reformulation of Article VI.  
Namely, the BSW action also has a bare 'square root of a homogeneous-quadratic kinetric term' factor, 
but this is broken by correction terms involving the shift $\u{\upbeta}$.

\m 

\n Manifestly Parametrization Irrelevant actions implement ii) explicitly. 
These moreover have the further benefit of being dual to geometrical actions: 
a viewpoint that does not even make reference what is irrelevant (i.e.\ parametrization).  
The current Article's final formulation of this second attribute is thus by use of geometrical actions. 
These involve change variables $\d \biQ$ rather than parameter-velocity ones.  

\m 

\n 2) A basic argument of Dirac establishes that Manifestly Reparametrization Invariant actions imply primary constraints. 
We uplift this argument to our preferred geometrical action setting. 
This means that Temporal Relationalism is a type of constraint provider, the next Article's Configurational Relationalism constituting another such. 
For Jacobi, BSW-type and Misner actions, homogeneous-quadraticity of the action leads to a constraint quadratic in the momenta.  
These examples' constraints are, respectively, the object elsewhere regarded as an energy constraint $\scE$, the famous GR Hamiltonian constraint $\scH$ 
and its minisuperspace restriction.  

\m 

\n 3) In the present setting, moreover, these are to be viewed as equations of time, and thus collectively denoted by $\Chronos$.
This interpretation refers to rearranging $\Chronos$ to implement Mach's Time Principle: that time is to be abstracted from change. 
This occurs, literally, via substituting the momentum-change relations 
-- the Temporally Relational replacement for momentum--velocity relations -- into $\Chronos$. 
This yields classical Machian emergent time. 

\m 

\n{\bf Remark 1} The title of BSW's paper, ``{\it Three-dimensional geometry as carrier of information about time}", supports the above duality.
Upon subsequently passing to the relational GR action, this can moreover be rephrased in the temporally Machian form 
`geometry and change of geometry as carrier of information about time'.   
GR's spatial geometries are moreover but an example of $\FrQ$ geometry. 
This can thus be further generalized as regards range of theories, to `{\sl Configuration and change of configuration as carrier of information about time}'.  

\m

\n 4) This is a major part of how the discrepancy between Leibnizian timelessness and us apparently experiencing time is resolved.
Namely, that Leibnizian timelessness, at the primary level, is Machianly mitigated at the secondary, i.e.\ emergent, level. 

\m 

\n 5) The other major part of this resolution follows from noticing that what we experience are subsystems, rather than the `universe as a whole'.  

\m 

\n{\bf Example 1} In the case of Mechanics, this classical Machian emergent time moreover simplifies the theory's equations down to coincide with Newton's.  
In this manner, we arrive at a recovery of what is `computationally, to good accuracy' Newtonian time, 
but now resting on relationally-acceptable foundations and possessing an emergent character. 
%
%
I.e.\ {\sl a relational recovery of Newtonian time}. 

\m 

\n The rotation of the Earth was long held to `read off' Newtonian time.  
This was subsequently found to be inaccurate to 1 part in $10^8$. 

\m 

\n A Machian view of this is as follows. 
The rotation of the Earth `reads off' classical emergent Machian time to the stated accuracy. 
The notion of replacing one subsystem by another to increase accuracy of time abstracted has moreover a solid grounding within the Machian perspective. 
Namely, one knows here to ask `which change' is `time' to be abstracted from, so that this `time' meets a required standard of accuracy.  
In the present example, for instance using the Earth--Moon--Sun system instead of just the Earth takes one past the stated bound on accuracy.  

\m 

\n 6) In general, as Chapter 15 of \cite{ABook} argues, `sufficient totality of locally relevant change (STLRC)' wins out over other authors' 
`any change' and `all change' positions on `which change'.  

\m 

\n The STLRC and `all change' positions moreover lie within the astronomers' {\it ephemeris time} conception \cite{Clemence}, 
in which other solar system bodies contribute to the timestandard. 
STLRC furthermore chooses further features of the ephemeris time conception this over `all change's' further Leibnizian-but-now-impractical tenets.  
In STLRC, all changes are given the opportunity to contribute, but only those found to be locally relevant are kept in the calculation. 
This is significant since, firstly, `all change' would include many poorly-measurable or even unobservable changes, 
Secondly, `all change' does not moreover translate well to features of relativisitic and quantum paradigms not anticipated by Leibnizian thinking 
(see Part I and Chapter 15 of \cite{ABook}).  
By this the time abstracted from STLRC merits the name {\it GLET: generalized local ephemeris time}. 

\m 

\n 7) This emergent time is {\sl provided by} the system; in this way, it complies with Mach's Time Principle.
In contrast, the notion of time usually assumed as an independent variable is neither Leibnizian nor Machian.   
Once the above time has been abstracted from change, it {\sl is} a convenient choice for (emergent) independent variable.  
A caveat on `system' here is that nature, not us, chooses what the system is. 
So e.g.\ if we wish to study a pendulum, our wish to study that pendulum has to include whatever else that pendulum cannot be isolated from, 
for intance that it is being studied in the terrestrial reference system. 
By which it is the Earth that is overwhelmingly locally relevant, and not the pendulum itself.

\m 

\n{\bf Remark 2} While nowadays time is read off atomic clocks, 
these are still to be interpreted as clock hands that are in regular need of calibration checks against solar system observations.  
Thus it is clear that there is a sense in which atomic clocks have not supplanted ephemeris time type concepts.  

\m 

\n This `reading hand' versus `calibration' distinction is moreover illustrative of how 
conceptual errors can arise from not separating out distinct notions within timekeeping. 

\m 

\n{\bf Remark 2} Whereas such an ephemeris time has long been in use, its Machian character has only relatively recently been remarked upon \cite{B94I, ARel2, ABook}.

\m

\n 8) Once armed with the above account, that quantum GR exhibits primary-level timelessness for the Universe as a whole is {\sl expected}, 
                                                                                                             rather than {\sl surprising}. 
This refers to the {\bf Frozen Formalism Problem} facet of the Problem of Time. 
Namely that the Wheeler--DeWitt equation \cite{Battelle, DeWitt67}  
\be 
\widehat{\scH} \Psi = 0 \m
\ee
-- the quantum version of the GR Hamiltonian constraint $\scH$, where $\Psi$ the waverfunction of the universe -- is, at least at first sight, 
a time-independent Schr\"{o}dinger equation arising at a juncture at which we are accustomed to seeing time-dependent Schr\"{o}dinger equations. 
This is now accounted for as an ab intio position resulting from the Temporal Relationalism aspect of Background Independence, 
now moreover preceded by awareness of how timelessness is already also ab initio classically present and then resolved in a Machian manner. 
While the classical emergent Machian time {\sl fails} to carry over to the quantum level, this is for a further Machian reason.
Namely that now {\sl quantum} change has to be given an opportunity to contribute, which can e.g.\ {\sl somewhat} change 
the form taken by the emergent Machian time in semiclassical models relative to that in the corresponding classical models.
Thus the same conceptualization carries over, but the form taken by the explicit Machian realization requires adjustment, 
in accord with the `GLET is to be abstracted from STLRC' perspective. 
See Article IV for brief further details, or Part III of \cite{ABook} for a full account.

\m 

\n{\bf Remark 3} Toward combined rather than piecemeal resolutions of facets, let us make our first distinction between 
    a Small Method -- for a piecemeal facet -- 
and a Large Method: suitable for combination to form A Local Resolution of the Problem of Time. 

\m 

\n Our Small Method is to use a Jacobi geometrical action (or its field-theoretic equivalent and/or generalization to a Jacobi--Synge action when required).  

\m

\n In contrast, our Large Method is to 
\be 
\mbox{\sl take Jacobi's action principle seriously enough to rederive the rest of Physics concordantly}. 
\label{Jac-Ser}
\ee 
Doing this carries a guarantee of {\sl remaining within} Temporal Relationalism as one succesively deals with each further local facet. 

\m 

\n This turns out to consist of the following. 

\m 

\n  1) Having to reconceptualize and rederive around half of the Principles of Dynamics material used, 
thus forming TRiPoD: the Temporal Relationalism implementing Principles of Dynamics. 
The other half turns out to be already-TRi, which is remarkably interesting given its strong overlap with successful approaches to constrained systems 
and with those parts of the Principles of Dynamics that are direct precursors of Quantum Theory's structures.  
This material is spread out over Articles I to X, and summarized in Fig XIII.6. 

\m 

\n 2) TRiFol: a TRi reformulation of foliation kinematics, as per Article XII. 

\m 

\n 3) TRiCQT (Canonical Quantum Theory) and TRiPIQT (Path Integral Quantum Theory); see Part III of \cite{ABook}.

\m 

\n The main virtue of the TRi formalism is that keeping one's calculations within this formalism prevents 
the Frozen Formalism Problem inadvertently re-entering while one is subsequently addressing further facets. 

\m 

\n The vast difference in size between our Small and Large methods is a reflection of how much more work it takes to resolve facet interferences 
as compared to just piecemeal facets.
For sure, the current Series provides Large Methods for all local facets of the Problem of Time at the classical level.  

\m 

\n{\Large\bf Acknowledgements for the whole Series \normalfont\normalsize}

\m

\n Foremost, I thank the people I am close with.
Without your support, I could have never written this.

\m 

\n For support when I was younger, I also thank in particular my father O and my friend L.

\m 

\n I thank Chris Isham in particular for discussions and support over the years in which I worked on this Series of Articles.  
I also thank my PhD supervisors Malcolm MacCallum and Reza Tavakol, 
and subsequently Enrique Alvarez, Jeremy Butterfield, Marc Lachi$\grave{\me}$ze-Rey and Don Page for support wth my career.  
For a range of discussions, comments, proof reading and thoughts, hosting, or support with my career, 
I also thank
Ozgur Acik,
Jeremy Butterfield, 
Malcolm MacCallum, 
Przemyslaw Malkiewicz, 
Don Page, 
Christopher Small, 
S and the others,   
Reza Tavakol,   
various participants at the Centenary meeting on Noether's Theorems, and numerous proofreaders.

\m 

\n Most of the thinking and calculating for this Series of Articles was done while at 
Peterhouse -- the University of Cambridge College -- as regards Articles I to VI, 
Universidad Autonoma de Madrid for Articles VII and VIII, 
Queen Mary: University of London for Article IX, 
Universit\'{e} Paris VII for Articles X to XIII, where I held a Foundational Questions Institute (fqXi.org). 
I also thank fqXi for a number of travel mini-grants.  
This work could have not been carried out if Cambridge's Moore Library (Mathematics) did not have 24/7 access, 
a matter in which Professor Stephen Hawking was pivotal.  
I generally encourage 24/7 access to academic libraries elsewhere as in the best interests of foundational research actually getting done.

\m 

\n I also thank my friends           $11A$, $3B$, $4C$, $D$, $4E$, $2F$, $G$, $2H$, $2J$, $2K$, $3L$, $3M$, $O$, $4R$, $7S$, $T$ and $W$
for keeping my spirits up at many points in this long journey. 

\m 

\n I finally wish to pay my respects to the above-mentioned Professor Stephen Hawking, 
as well as to Professor John Stewart, who had strongly encouraged my study of Lie derivatives.


\end{document}